\documentclass{article}

\RequirePackage[OT1]{fontenc}
\RequirePackage{amsthm,amsmath}
\usepackage{amsfonts}

\usepackage[applemac]{inputenc} 
\usepackage{graphicx}
\usepackage{color}
\usepackage{amsfonts}
\usepackage{changes}

\begin{document}

 \large

\newcommand{\al}{\mbox{$\alpha$}}
\newcommand{\be}{\mbox{$\beta$}}
\newcommand{\ep}{\mbox{$\epsilon$}}
\newcommand{\gam}{\mbox{$\gamma$}}
\newcommand{\sig}{\mbox{$\sigma$}}

\DeclareRobustCommand{\FIN}{%
  \ifmmode 
  \else \leavevmode\unskip\penalty9999 \hbox{}\nobreak\hfill
  \fi
  $\bullet$ \medskip}

\newcommand{\calA}{\mbox{${\cal A}$}}
\newcommand{\calB}{\mbox{${\cal B}$}}
\newcommand{\calC}{\mbox{${\cal C}$}}

\newcommand{\muas}{\mbox{$\mu$-a.s.}}
\newcommand{\Nat}{\mbox{$\mathbb{N}$}}
\newcommand{\Rea}{\mbox{$\mathbb{R}$}}
\newcommand{\Prob}{\mbox{$\mathbf{P}$}}

\newcommand{\nin}{\mbox{$n \in \mathbb{N}$}}
\newcommand{\suc}{\mbox{$\{X_{n}\}$}}
\newcommand{\sucP}{\mbox{$\mathbf{P}_{n}\}$}}

\newcommand{\Pd}{\mbox{$\mathcal{P}_2(\mathbb{R}^d)$} }
\newcommand{\subPd}{\mbox{$_{\mathcal{P}_2(\mathbb{R}^d)}$} }
\newcommand{\Wd}{\mbox{$\mathcal{W}_2$}}
\newcommand{\Wdd}{\mbox{$\mathcal{W}_2^2$}}
\newcommand{\Read}{\mbox{$ \mathbb{R}^d$} }
\newcommand{\Exp}{\mbox{$ \mathbf{E}$}}
\newcommand{\muo}{\mbox{$ \mu_\omega $}}

\newcommand{\Aal}{\mbox{$ {\cal T}^\alpha_\mathbf{P} $}}
\newcommand{\calX}{\mbox{$ {\cal X}$}}
\newcommand{\calQ}{\mbox{$ {\cal Q}$}}
\newcommand{\calQX}{\mbox{$ {\cal Q}^{\cal X}$}}
\newcommand{\cBall}{\mbox{$\overline{B}_\mathcal{W}($}}
\newcommand{\nuP}{\mbox{$\nu_{\mathbf{P}}$}}
\newcommand{\nuPn}{\mbox{$\nu_n^\omega$}}

\newcommand{\conv}{\rightarrow}
\newcommand{\convn}{\rightarrow_{n\rightarrow \infty}}
\newcommand{\convp}{\rightarrow_{\mbox{c.p.}}}
\newcommand{\convs}{\rightarrow_{\mbox{a.s.}}}
\newcommand{\convw}{\rightarrow_w}
\newcommand{\convd}{\stackrel{\cal D}{\rightarrow}}

\newtheorem {Prop}{Proposition} [section]
 \newtheorem {Lemm}[Prop] {Lemma}
 \newtheorem {Theo}[Prop]{Theorem}
 \newtheorem {Coro}[Prop] {Corollary}
 \newtheorem {Nota}[Prop]{Remark}
 \newtheorem {Ejem}[Prop] {Example}
 \newtheorem {Defi}[Prop]{Definition}
 \newtheorem {Figu}[Prop]{Figure}
 \newtheorem {Tabla}[Prop]{Table}

\title{\sc  Wide  Consensus  aggregation in the Wasserstein Space. Application to location-scatter families.\footnote{Research partially supported by the
Spanish Ministerio de Econom\'{\i}a y Competitividad y fondos FEDER, grants  
MTM2014-56235-C2-1-P, and MTM2014-56235-C2-2 , and  by Consejer\'{\i}a de Educaci\'on de la Junta de Castilla y Le\'on, grant VA212U13.}}

\author{Pedro C. \'Alvarez-Esteban$^{1}$, E. del Barrio$^{1}$, J.A. Cuesta-Albertos$^{2}$\\ and C. Matr\'an$^{1}$ \\
$^{1}$\textit{Departamento de Estad\'{\i}stica e Investigaci\'on Operativa and IMUVA,}\\
\textit{Universidad de Valladolid. SPAIN} \\ $^{2}$ \textit{Departamento de
Matem\'{a}ticas, Estad\'{\i}stica y Computaci\'{o}n,}\\
\textit{Universidad de Cantabria. SPAIN}}
\maketitle

\begin{abstract}
We introduce a general theory for a consensus-based combination of estimations of probability measures. 
Potential applications include parallelized or distributed sampling schemes 
as well as variations on aggregation from resampling techniques like  boosting 
or bagging. Taking into account the possibility of very discrepant estimations, 
instead of a full consensus we consider a ``wide consensus" procedure. 
The approach is based on the consideration of trimmed barycenters in the Wasserstein 
space of probability measures. We provide general existence and  consistency results
as well as suitable properties of these robustified Fréchet means{. In order to get quick applicability, we also} include characterizations of barycenters of probabilities that belong to (non necessarily elliptical)  
location and scatter families. For these families we provide an  iterative algorithm for 
the effective computation of trimmed barycenters,  based on a consistent algorithm for computing barycenters, guarantying applicability in a wide setting of statistical problems.
\end{abstract}

\noindent {\small \textsc{AMS Subject Classification:} Primary:  60B05, 62F35, Secondary 62H12. }

\noindent {\small \textsc{Keywords:} Trimmed barycenter, wide consensus, robust aggregation, Wasserstein distance, trimmed distributions, impartial trimming, parallelized inference.}

\section{Introduction.}

Data that consists of samples composed by probability distributions are increasingly common. 
Examples include the distribution of a set of medical measurements in hospitals in a multicenter clinical
trial or that of several economic magnitudes (income and age distribution, for instance) in different countries.
Often these distributions are not directly observed, but some estimation is available. This paper 
introduces a new approach for the combination of several estimations of probabilities. Our goal is to provide a tool to 
combine available estimations to get a consensus-based global estimation. 
We recall that this goal has been largely pursued under different frameworks. 
Merging  information, pooling estimation, aggregation estimation or meta-analysis, 
are expressions related with this common  goal. The potential applications that we have in mind also
include parallelized or distributed estimation schemes as well as those provided by 
resampling methods designed to improve unstable procedures or to provide approximate 
solutions through algorithms involving combinatorial complexity problems.

At present, statistical methodologies under a parallelized or distributed scheme are 
receiving growing interest. In fact, they constitute a basic statistical challenge in a 
world where we want to exploit massive data sets that could 
have been collected by different units or that exceed the size that would make their analysis on a single machine feasible. 
The need for aggregation methods becomes clear in the following two cases. One, when  the different sets of data would be obtained, stored and even processed by the different units, 
perhaps using different experimental techniques. Another, associated to the ``divide and conquer" 
principle, would include the combination of results obtained from the partition of the data set in smaller, 
tractable subsets.  Note that the partition of the data, in this second category, is often performed 
{based on computational convenience criterions}, say by their storage {location},  or oldness in 
the data basis, hence essentially both categories share the same handicap: the hypothesis of 
homogeneity of the distributions corresponding to the different units seems to be excessively optimistic in practice.

Regarding the already mentioned resampling methods, since the introduction of bagging by Breiman \cite{Breiman}, 
subagging,  and other aggregating procedures have been introduced in the last years to improve the performance of estimators in different setups, including regression
or classification (see e.g. 
B\"uhlmann and Yu \cite{Buhlmann2}, B\"uhlmann \cite{Buhlmann3} and B\"uhlmann and Meinshausen \cite{Buhlmann}). The aggregation  is usually achieved  just by averaging, but there are also other proposals like bragging 
(in \cite{Buhlmann3})) or magging (in \cite{Buhlmann}), which aim at robust aggregation.
In a different problem, the available algorithms for the obtention of some well known estimators 
(like the  Minimum Volume Ellipsoid (MVE), Minimum Covariance Determinant (MCD) and several others) 
involve the use of a iterative procedure starting from many initial random choices, either for statistical or
computational convenience, that result in a set of different estimations that must be combined to produce
a better (or just computable) estimation (see e.g. Woodruff and Rocke \cite{Woodruff}, Croux and Haesbroeck \cite{Croux}, 
or Rousseeuw and Driessen \cite{RousseeuwDriessen}). Depending on the intrinsic geometry of the estimated objects, 
the aggregation procedure may have to be based on some sort of non standard averaging technique.

Aggregation of a set of estimations of probabilities to provide a final estimation -- the consensus -- 
is analyzed in this paper under a novel point of view. We can get motivation for our goal from
the following hypothetical situation. Consider a biomedical study to be carried out and processed by a network of hospitals. 
Each hospital will provide an estimation of the distribution of interest and the goal is to obtain a meta-estimation summarizing 
the  estimations. This combination of information is sensitive to two different possible types of atypic or noisy 
data. First, the sample obtained in any hospital could have some contaminating data. Second, one or several hospitals 
could produce very atypical results when compared to the others simply because the patients in the influence zone of the 
hospital have very different { (social, cultural, ethnic, nutritional) features}.
To handle this general setting we will assume that there exist $k$ units, say $U_1,\dots,U_k$, and that unit $U_i$ 
will process a sample $x_1^i, ...,x_{n_i}^i$ of $\Read-$valued data obtained from a distribution $P_i$. As 
the results of processing their associated samples, the units produce a new sample 
consisting in the estimations $\hat P_1,\dots,\hat P_k$, perhaps given through the estimations of suitable parameters.  
Our goal will be to produce a consensus estimator from  those obtained by the different units. 
However, since some units could process very contaminated batches, whose consideration would lead to 
large deviations {from the mainstream model} (if any), we will include the possibility of obtaining a 
wide consensus instead of a full consensus. 
In our scheme the meaning of wide consensus must be understood as the possibility of 
avoiding the results of the most discrepant units, elaborating  the consensus just from the {remaining units}.

We emphasize that our approach assigns a different status to the samples processed by the units  
(composed by points in $\Rea^d$) and to the  meta-sample, of size $k$  
(composed by probability distributions on $\Rea^d$), provided by the units. 
The primitive samples have the usual meaning in Statistics and will be processed through more or less 
standard procedures, a task that will not be considered here,  
our object of interest being the sample of probability distributions. To work with
 this sample we will make a careful use of the structure of these objects. To illustrate this point,
let us consider a simple example involving estimation in a normal model. Our proposal aims  
at producing a normal distribution which is an optimal representation of $k$ normal distributions, 
$\hat P_1,\dots,\hat P_k$ in some sense. Note that a (weighted) average of probabilities is a probability, 
but the mixture of normal distributions that we could produce in this way is not normal and could be very far, in terms of shape,
from the $k$ normal distributions that we are trying to summarize, hence making  a different aggregation procedure 
to be more convenient.

Our choice for the basic aggregation procedure is the Wasserstein barycenter, that we briefly describe next.
We will work in the space $\Pd$ of probability measures on $\Read$  with 
finite second order moment, endowed with the $L_2$-{\it Wasserstein  distance}, $\mathcal{W}_2$, defined for $P,Q\in \Pd$ by 
\begin{equation} \label{int1}
\Wd (P,Q):=\inf \left\{\left(\Exp\|U-V\|^2\right)^{1/2}: \  \mathcal{L}(U)=P, \  \mathcal{ L}(V)=Q\right\},
\end{equation}
where we use $\mathcal{L}(X)$ to denote the distribution law of a r.v. $X$. 
Given a finite set of elements, $P_1,\ldots,P_k\in \Pd$ and positive weights $\lambda_1,\ldots,\lambda_k$, with $\sum_{i=1}^k\lambda_i=1$, we would like 
to obtain a representative element for the whole set. Like the mean of a set of vectors, a barycenter or Fr\'echet mean in this space can be a good candidate and would be any probability, $\bar P\in \Pd$ { satisfying}
\begin{equation}\label{barycenter}
\sum_{i=1}^k\lambda_i \mathcal{W}_2^2(\bar P,P_i) = \inf\left\{\sum_{i=1}^k\lambda_i \mathcal{W}_2^2(P,P_i): \ \  P\in \Pd\right\}.
\end{equation}
Such a probability, when it exists, is called a {\it ($\{\lambda_i\}_{i=1}^k-$weighted) barycenter} of $\{P_i\}_{i=1}^k$. The 
consideration of barycenters in the Wasserstein setting has been initiated by Agueh and Carlier in \cite{Agueh10}, with several extensions in Boissard et al \cite{Bois15}, Pass \cite{Pass}, Bigot and Klein \cite{Bigot} and in Le Gouic and Loubes \cite{Le Gouic}, where the concept has been extended to  arbitrary (non-necessarily finite) families of probabilities (see Definition \ref{family} below). 

A {\it full consensus representation} of $P_1,\ldots,P_k$ would be the barycenter associated to equal weights $\lambda_i=1/k, i=1,\dots,k$. 
Different weights would be more appropriate if, for example, some of the $P_i$'s have been obtained from (or represents) a considerably larger population than some others.  
{ On the other hand, the possible} existence in $P_1,\ldots,P_k$ of very discrepant representations, 
(possibly due to highly contaminated batches as before), would justify a trimming or reweighting action. 
Rather than using the ($\{\lambda_i\}_{i=1}^k-$weighted) barycenter of $\{P_i\}_{i=1}^k$, with the original weights, 
the {\it wide consensus representation} of  $P_1,\ldots,P_k$ with weights $\{\lambda_i\}_{i=1}^k$ or {$\alpha$-trimmed barycenter}, $\bar P_\alpha$, 
is a solution, for suitable weights $(\bar \lambda_i^{\alpha})_{i=1}^k$, of the following double minimization problem
\begin{equation}\label{trimbar}
\sum_{i=1}^k\bar \lambda_i^{\alpha} \mathcal{W}_2^2(\bar P^{\alpha},P_i) = \inf\Big\{\sum_{i=1}^k \lambda_i^* \mathcal{W}_2^2(P,P_i): \ 
P\in \Pd, \  \lambda_i^*\leq \lambda_i, \ \sum_{i=1}^k \lambda_i^* = 1 - \alpha\Big\}.
\end{equation}

Trimming procedures are of frequent use in Robust Statistics to { prevent} the influence of atypical data in statistical analyses. In fact, a trimmed version of the Wasserstein distance for probabilities on the line was introduced, in the context of Goodness of Fit tests, by Munk and Czado \cite{Munk} to  avoid the effects of data in the tails. This approach was extended in some papers (see Álvarez-Esteban et al \cite{Alvarez-Esteban2012} and references therein) to cover trimmings like that considered in (\ref{trimbar}) that are
``impartial". This means  that there are not  a priori selected directions or zones for trimming, 
being the complete data set which will provide that information. 
Although often trimming is used with the meaning of deleting a part of the data, 
here we follow a more flexible approach as in Gordaliza \cite{Gordaliza}, based on
probability trimmings (see Definition \ref{definition1} below) which allows to decrease 
the weight of some regions without completely remo\-ving them. 
We include in subsection \ref{overview} a succinct account of basic results on probability trimmings 
and refer to \'Alvarez-Esteban et al \cite{Alvarez-Esteban2011} for further details.

In this paper we introduce the concept of trimmed barycenter for probabilities $\mu$ on the Wasserstein space of probabilities on $\mathbb{R}^d$ 
with finite second moment endowed with the metric $\mathcal{W}_2$, extending that of
trimmed mean introduced in Rousseeuw \cite{Rousseeuw} and Gordaliza \cite{Gordaliza}. Notice that no moment assumption is made on $\mu$.
Our setup covers the case of general Borel probability measures on  $\Pd$ (of which (\ref{trimbar})
corresponds to the particular case of finitely supported measures, see Definitions \ref{family} and \ref{definition1} below).
We provide existence and consistency results for trimmed barycenters.  {In particular we prove a Strong Law of Large Numbers (Theorem \ref{consistency2}) for trimmed barycenters in this space, to be denoted throughout by
$W_2(\Pd)$ (see (\ref{W2space}) below). }

As noted before, a desirable feature of any aggregation method is  adaptation to the
shape of the objects to be aggregated. Remarkably, this is the case for barycenters and trimmed
barycenters in location and scatter families 
such as the Gaussian family. In particular, we show that the barycenter or trimmed barycenter 
of a probability on $\Pd$ supported in a (non-necessarily finite) set of  probabilities belonging to a location scatter family also belongs to the family.
We also provide a characterization of barycenters in location and scatter families in terms of a 
fixed point equation as well as some equivariance results for general 
barycenters and trimmed barycenters. Notice that suitability of the location and scatter families in the Wasserstein space has been  considered by Chernozhukov et al. in \cite{Depth} in relation with Monge-Kantorovich quantiles.  Also,  Rippl et al. \cite{Rippl} take advantage of the explicit expression of the Wasserstein distance between Gaussian distributions. In a similar spirit to that considered here, they substitute sampling distributions obtained from Gaussian distributions by Gaussian distributions with estimated parameters, and address the problem of the asymptotic behavior of the Wasserstein distance between empirical and theoretical distributions. Their analysis includes the two-sample setting, for independent samples, through the distance between normal distributions when the parameters are estimated from the respective samples.

Turning back to the statistical motivation of this work, we note that the applicability of
barycenters or trimmed barycenters for data analysis will strongly depend on the availability
of efficient algorithms for their computation. In this sense, we stress the fact that, in the 
multivariate setting, even the computation of the barycenter
of a finite collection of normal distributions can be a hard task since no closed form expression
for the barycenter is available. On the other hand, convexity of the map $\eta\mapsto \mathcal{W}_2^2 (\eta,\nu)$
implies that Wasserstein barycenters are minimizers of a convex functional. This fact is
at the basis of a fast algorithm just introduced in \cite{preprint} for the approximate computation
of barycenters, including the case of location and scatter families. 
Here we show how this can be used for the efficient computation of trimmed barycenters in these
location and scatter families. {Also we must stress that our approach constitutes a technical keystone for the introduction of robust clustering in the Wasserstein space, opening new applications in that wide setting (see del Barrio et al. \cite{preprint3}).}

The remaining sections of this paper are organized as follows. In Section \ref{preliminary} we give a quick
account of notable results on Wasserstein distance and Wasserstein spaces including the main known results on 
barycenters. Section \ref{TrimmedBar} introduces trimmed barycenters and provides the main results announced before. They
include existence, consistency,  equivariance, and characterizations in  location-scatter families as well as relevant properties involving shapes and sizes.
Section \ref{computation} discusses computational issues for barycenters and trimmed barycenters
and presents an algorithm for the computation of trimmed barycenters in location
and scatter families. It also includes some toy examples and an application  to aggregation of MCD's solutions obtained by subsampling on a real data set. The analysis of this example includes hints on the possible selection of the trimming as well as information on the running times of the algorithms.  Most of the technical details and proofs are deferred to Section \ref{appendix}.

{
We conclude this Introduction with some explanations on notation. 
Unless explicitly noted, probability measures are defined on the Borel 
$\sigma$-algebra of the (metric) space. The indicator function of a set, 
$A$, will be represented by $I_A$, while $\delta_{\{x\}}$ will denote Dirac's
measure on $x$. We write $\ell_d$ for Lebesgue measure on 
$\mathbb{R}^d$ and $\mu\ll\nu$ to mean that $\mu$ is absolutely continuous
with respect to $\nu$. Weak convergence of probability measures will be denoted by $\convw$. 
We assume that weights, $\lambda_1,\dots,\lambda_k$, are positive numbers, $\lambda_i>0, i=1,\dots,k$ such that $\sum_{i=1}^k\lambda_i=1.$
We will denote by $\mathcal{P}_{2,ac}(\Read)$ the subset of  absolutely continuous probabilities (with respect to $\ell_d$) 
in $\Pd$. Given $P \in \Pd$ and $r>0$,  $B_{\mathcal{W}} ( P,r)$ (resp. $\cBall P,r)$) will be 
the open (resp. closed) ball with center at $P$ and radius $r$ for the distance  $\Wd$, while $B(x,r)$, where $x \in \Rea^d$, will refer 
to the open  ball  with center at $x $ and radius $r$ for the Euclidean distance on $\Rea^d$. 
Finally, we will say that the map $T$ {\it transports} (pushes forward) the probability 
$P$ to $Q$ if $Q$ is the image measure of $P$ by $T$, namely, if  $Q=P\circ T^{-1}$.
}

\section{ Barycenters in Wasserstein space} \label{preliminary}

As noted in the Introduction, our proposal for wide consensus aggregation is based on
Wasserstein metrics and barycenters in Wasserstein space.
We refer to the books of Villani \cite{Villani}, \cite{Villani2} for a complete and 
 well documented view of the general theory on Wasserstein spaces and optimal 
transport and to the papers by Agueh and Carlier \cite{Agueh10} and Le Gouic and Loubes 
\cite{Le Gouic} for barycenters. Here we include a brief introduction, continued in Subsection \ref{WassSpaces},
with some relevant facts and necessary results for our presentation. 

It is well known  that   the infimum in (\ref{int1}) is attained, i.e., there exists a pair $(X,Y)$, defined on some probability space, 
with ${\cal L}(X)=P$ and
${\cal L}(Y)=Q$ such that $\Exp\|X-Y\|^2=\mathcal{W}_2^2(P,Q).$
Such a pair $(X,Y)$ is called a {\it \Wd-optimal transportation plan} (\Wd-o.t.p.)  
for $(P,Q)$, although the alternative terminology {\it $L_2$-optimal coupling} for $(P,Q)$  is often used.

For probabilities on the real line, it is well known that the quantile functions associated to $P$ and $Q$, 
denote them by $F_P^{-1}$ and $F_Q^{-1}$, are a \Wd-o.t.p.,
\begin{equation}\label{casoreal}
\Wd(P,Q)=\Big(\int_0^1\big(F_P^{-1}(t)-F_Q^{-1}(t)\big)^2dt\Big)^{1/2},
\end{equation}
but for multivariate distributions there is no equivalent explicit expression to compute $\Wd(P,Q)$. 
A useful fact, that allows to focus on the case of centered probabilities 
is that if $m_P,m_Q$ are the means of $P$ and $Q$, and $P^*,Q^*$ are the corresponding centered probabilities, then
\[
\Wdd(P,Q)=\|m_P-m_Q\|^2+\Wdd(P^*,Q^*).
\]

{ In the late 1980s and early 1990s, a series of papers by Brenier \cite{Brenier1,Brenier2}, Cuesta-Albertos and Matr\'an
\cite{Cuesta} and R\"uschendorf and Rachev \cite{Ruschen} put the basis for the analysis of optimal transporting: under continuity assumptions on
the probability $P$, the $L_2$-o.t.p. $(X,Y)$ for $(P,Q)$
can be represented as $(X,T(X))$ for some suitable 
map $T$. Moreover, }this {\it optimal transport map}  for $(P,Q)$  coincides with the (essentially unique) cyclically monotone map transporting $P$ to $Q$.

A very interesting consequence of the characterization of optimal transportation maps is 
that, independently of the initial distribution, some maps have the optimal transport 
property between any initial probability $P$ and its transported probability. 
In particular, if $A$ is a positive definite matrix (here and through the 
paper we assume that positive definiteness  includes symmetry), then $(X,AX)$ 
is a \Wd-o.t.p. independently of the law $\mathcal{L}(X)$. This 
fact allows to characterize the optimal transport maps between nonsingular normal 
distributions and yields some additional facts that we quote in the next result, 
a version of  Theorem 2.1 in Cuesta-Albertos et al. \cite{Cuesta96}, which, in turn, 
improves the original statement by Gelbrich \cite{Gelbrich}. 

\begin{Theo}\label{Gelbrich2}
Let $P$ and $Q$ be probabilities in \Pd with means $m_P, m_Q$ and  covariance matrices $\Sigma_P, \Sigma_Q$. If $\Sigma_P$ is assumed nonsingular, then 
\begin{eqnarray}
\nonumber
\Wdd(P,Q)&\geq& \|m_P-m_Q\|^2+ trace\left(\Sigma_P+\Sigma_Q-2\left(\Sigma_P^{1/2}\Sigma_Q\Sigma_P^{1/2}\right)^{1/2}\right)
\\
\label{cotaGelbrich}
 &=& \Wdd(N(m_P,\Sigma_P),N(m_Q,\Sigma_Q)).
\end{eqnarray}
Moreover the equality holds if and only if the map $T(x)=(m_Q-m_P)+Ax$ transports $P$ to $Q$ (in particular if $P$ and $Q$ are gaussian), where $A$, semidefinite positive, is defined by 
\begin{equation}\label{transportenormales}
A:=\Sigma_P^{-1/2}\left(\Sigma_P^{1/2}\Sigma_Q\Sigma_P^{1/2}\right)^{1/2}\Sigma_P^{-1/2}, 
\end{equation}
\end{Theo}

The set \Pd equipped with the \Wd-distance is a Polish space (separable and complete metric space) that is often 
called a {\it Wasserstein space} and denoted as $W_2(\Read)$. We can also consider 
(through a definition of the distance similar to that in (\ref{int1})) a Wasserstein-type space over other spaces,
notably over \Pd leading to $W_2(\Pd)$. This space consists of the probability measures, $\mu$, on \Pd (equipped 
with the Borel $\sigma$-field associated to the distance \Wd) such that 
\begin{equation}\label{W2space}
\int_{\Pd}\mathcal{W}_2^2(P,Q)\mu(dP) < \infty, \ \mbox{ for some (hence, for every) } Q \in \Pd.
\end{equation}
Wasserstein distance in this space will be denoted by $\mathcal{W}_{\mathcal{P}_2}$.
{It is worthwhile to stress that the Wasserstein metric on $W_2(\Pd)$ inherits the { good} properties that it exhibits on $\Pd$ (see subsection \ref{WassSpaces}).
The space $W_2(\Pd)$ is in the basis of the  (more abstract) framework considered in \cite{Le Gouic} to generalize (\ref{barycenter}) to this definition of barycenters. 

\begin{Defi}\label{family}
If $\mu \in W_2(\Pd),$ then  a barycenter of $\mu$ is any probability $\bar \mu \in \Pd$ such that $\mathcal{W}_{\mathcal{P}_2}^2(\mu,\delta_{\{\bar \mu\}})=\inf\{\mathcal{W}_{\mathcal{P}_2}^2(\mu,\delta_{\{Q\}}), Q\in \Pd\}$, that is:
\begin{equation}\label{family2}
\int_{\Pd}\mathcal{W}_2^2(P,\bar \mu)\mu(dP) = \mbox{\em Var}(\mu):= \inf \left\{\int_{\Pd}\mathcal{W}_2^2(P,Q)\mu(dP): Q \in \Pd\right\}
\end{equation}
\end{Defi}
We use the notation $\mbox{Var}(\mu)$ to stress the role of variance of $\mu$ played by this quantity.
Note that (\ref{family2}) is the natural extension of the already considered barycenters of a finite set of probabilities $P_1,\ldots,P_k \in \Pd$ with weights $\lambda_1,\ldots, \lambda_k$.
}

It will be convenient to consider a generic probability space $(\Omega,\sigma, \Prob)$ where a measurable 
random element with values in $\Pd$ (and distribution law $\mu$) is defined. The 
image of a generic $\omega\in\Omega$ will be denoted as $\mu_\omega$. Then equation (\ref{family2})
becomes
\begin{equation}\label{notacionbar}
\int_{\Omega}\mathcal{W}_2^2(\mu_\omega,\bar \mu)\Prob(d\omega) = \inf \left\{\int_{\Omega}\mathcal{W}_2^2(\mu_\omega,Q)\Prob(d\omega): Q \in \Pd\right\}.
\end{equation}
Existence of barycenters in this setting has been proved in \cite{Le Gouic}, as well as
uniqueness under absolute continuity assumptions (in fact this follows easily from 
Theorem 2.9  in  \cite{Alvarez-Esteban2011}). Barycenters in Wasserstein
space enjoy some continuity properties. We refer to Proposition \ref{existuni} and 
Theorems \ref{consistencyLoubes} and \ref{consistency} (which are essentially contained in Theorems 2 and 3 in \cite{Le Gouic})

We show next that barycenters in Wasserstein space satisfy an equivariance property with respect to 
similarity transformations, namely, linear transformations that preserve shape. 
We recall that these transformations include rotations, reflections, translations and scaling. 
A proof can be found in subsection \ref{Sec.ProofProposition}.
 
\begin{Prop}\label{equiv}
Let $\bar \mu \in \Pd$ be a barycenter of $\mu \in W_2(\Pd),$ and let $T$ be 
a similarity transformation on $\Rea^d$. If $\mu^*$ is defined as the probability 
in $W_2(\Pd)$ given, through the notation above, by $\mu^*_\omega=\mu_\omega\circ T^{-1}$, 
then $\bar \mu \circ T^{-1}$ is a barycenter of $\mu^*$.
\end{Prop}

We close this section with some remarks on the computability of Wasserstein barycenters.
In general, it shares the serious computational difficulties inherent to optimal transportation. 
Explicit expressions are available just for distributions on the real line, a fact that is
quoted in the next result.

\begin{Prop}\label{casoreal2}
 If $F_1^{-1},\ldots,F_k^{-1}$ are the quantile functions associated to probabilities $P_1,\ldots,P_k$ on the real line, and $\lambda_1,\ldots,\lambda_k$ are positive weights with $\sum_{i=1}^k\lambda_i=1$, then the barycenter of $\{P_i\}_{i=1}^k$ is the probability with quantile function $\sum_{i=1}^k\lambda_iF_i^{-1}$. 
\end{Prop}

From Proposition \ref{casoreal2} we see that for $k$ normal distributions, $N(m_i,\sigma_i^2), i=1,\dots,k,$, on \Rea, 
the barycenter would be the normal law $N(\sum_{i=1}^k\lambda_im_i,(\sum_{i=1}^k\lambda_i\sigma_i)^2)$. 
More generally, for multivariate normal distributions there is an interesting 
characterization for the barycenter that comes from Knott and Smith \cite{Knott} (but see also Rüschendorf and Uckelmann \cite{Ruschen2} and \cite{Agueh10}).

\begin{Theo}\label{casonormal}
Let $P_i=N(m_i,\Sigma_i), i=1,\ldots,k$ be normal probabilities 
on \Read with positive definite covariances, and $\lambda_1,\ldots,\lambda_k$ positive 
weights with $\sum_{i=1}^k\lambda_i=1$. Then the unique barycenter of $P_1,\ldots,P_k$ is the normal law $N(\bar{\mu},\bar{\Sigma})$,
where $\bar m=\sum_{i=1}^k\lambda_im_i$ and $\bar\Sigma$ is the only positive definite root of the equation
\begin{equation}\label{ecuacion}
\sum_{i=1}^k\lambda_i\left(\Sigma^{1/2}\Sigma_i\Sigma^{1/2}\right)^{1/2}=\Sigma.
\end{equation}
\end{Theo}

Later, in Theorem \ref{self-suf} we will generalize this result to probabilities in $W_2(\Pd)$ 
supported in an arbitrary location-scatter family. We note that our proof is elementary and self-contained
(in particular, it does not use Theorem \ref{casonormal} but only general principles of optimal transportation).

\section{ Trimmed Barycenters}\label{TrimmedBar}

We introduce in this section our approach for a wide consensus representative of a 
sample $P_1,\dots,P_k$ of probabilities, with given weights $\lambda_1,\dots,\lambda_k$.
It is based on considering a suitably trimmed subsample. The trimming procedure  allows 
partial discarding of some probabilities, through a suitable reweigthing as in the following definition.
\begin{Defi} \label{definition1} Given $0\leq\alpha\leq 1$ and $P$ a probability on a measurable space $(\Omega,\sigma)$, we say that the probability
$P^*$, also defined on $\sigma$, is an $\alpha$-trimming of $P$ if there exists a measurable function $\tau:\Omega \to \Rea$ such that $0\leq \tau(\omega) \leq 1$ for every $\omega \in \Omega$ and $P^*(A) =\frac 1 {1-\alpha} \int_A \tau(\omega)P(d\omega)$ for every $A\in \sigma$. {Such a} function is often called an $\alpha$-trimming function. The set of all $\alpha$-trimmings of $P$  will be denoted by ${\cal T}_\alpha (P)$
\end{Defi}

\begin{Nota}{\rm
A typical trimming function would be the indicator function of a set $A$ with probability $P(A)=1-\alpha$. The trimmed probability being then  the conditional probability given $A$. However, our definition even includes the consideration of $P$, itself, as a trimmed version of $P$, with associated trimming function $\tau=(1-\alpha)I_\Omega.$

Since trimmed probabilities and trimming functions are associated in an essentially one to one way, the notation ${\cal T}_\alpha (P)$ will be indistinctly used for the set of all $\alpha$-trimmings  of $P$  and for the set of the corresponding trimming functions.
}
\end{Nota}
Given $\alpha \in (0,1)$, and a probability $\mu$ on $\Pd$, we look for a  $\bar \mu^\alpha \in {\cal P}_2(\Rea^d)$ and a probability $\mu^\alpha \in {\cal T}_\alpha(\mu)$, with associated trimming function $\tau_{\mu}^\alpha$, which satisfy
\begin{equation}\label{def1trimbar}
\int  \Wdd(P, \bar \mu^\alpha) \mu^\alpha (d P ) = \mbox{Var}_\alpha(\mu):=\inf_{\mu^* \in {\cal T}_\alpha(\mu), \nu \in  {\cal P}_2(\mathbb{R}^d)} \int {\cal W}_2^2 (P, \nu) \mu^* (d P ) 
\end{equation}
or, { equivalently, in terms of trimming functions},
\[
\int  \Wdd(P, \bar \mu^\alpha)\tau_{\mu}^\alpha(P) \mu (d P ) 
=  (1-\alpha)\mbox{Var}_\alpha(\mu)= 
\inf_{\tau \in {\cal T}_\alpha(\mu), \nu \in  {\cal P}_2(\mathbb{R}^d)} \int {\cal W}_2^2 (P, \nu) \tau(P) \mu (d P ) .
\]
Such a $\bar \mu^\alpha$  will be called {\it ($\alpha$-)trimmed barycenter of $\mu$} and $\tau_{\mu}^\alpha$ 
an {\it ($\alpha$-)optimal trimming function}. Similarly to Var$(\mu)$, the  value $\mbox{Var}_\alpha(\mu)$  
will be called the {\it ($\alpha$-)trimmed variance} of $\mu$. 
As usually, the previous definitions apply to any $\Pd$-valued random variable, by 
identifying these concepts for  a random variable with those of its probability distribution. 
The following theorem (proved  in subsection \ref{Sec.Consistency}) guarantees the existence of trimmed barycenters.

\begin{Theo} \label{Theo.Bary}
Let $\alpha \in (0,1)$ and let $\mu$ be a probability defined on $\Pd$. Then, there exists an $\alpha$-trimmed barycenter of $\mu$, which we will denote as $\bar \mu^\alpha$.
\end{Theo}

By considering as trimming function  (with the corresponding normalizing factor), the indicator 
set of a large enough ball centered at $\delta_{\{0\}}$, it becomes obvious that the minimum 
value $\mbox{Var}_\alpha(\mu)$ must be finite and (recall the definition of $W_2(\Pd)$ in 
(\ref{W2space})) that the set  ${\cal T}_\alpha(\mu)$ can be substituted by the subset 
${\cal T}_\alpha(\mu)\cap W_2(\Pd)$. Since every probability on $W_2(\Pd)$ has a barycenter, 
obviously $\bar\mu^\alpha$ must be a barycenter of $\mu^\alpha$, which justifies the notation we are using. 
Furthermore, and similar to the  impartially trimmed means, trimmed barycenters must simultaneously be
the barycenter of the trimmed distribution and the center    of its support. 
To formalize this fact we define
\begin{equation}\label{radios}
r_\alpha (P) := \inf \{ r>0: \mu [B_{\mathcal{W}}(P,r)] \geq 1- \alpha \}.
\end{equation}
It trivially follows that if $r < r_\alpha (P)$, then $\mu [B_{\mathcal{W}}(P,r)] < 1 -\alpha$ and
\[
\mu [B_{\mathcal{W}} (P,r_\alpha (P)) ] \leq 1- \alpha \leq \mu [ \cBall P,r_\alpha (P))].
\]
This is the key to the following result.
 
\begin{Prop} \label{Lemm.charac2}
Let $\alpha \in (0,1)$, $\nu \in {\cal P}_2(\Rea^d)$ and $\tau^* \in {\cal T}_\alpha(\mu)$ be such that
\begin{equation} \label{Eq.2v.1}
I_{B_\mathcal{W} (\nu,r_\alpha (\nu) )} \leq  {\tau^*}  \leq I_{  \overline{B}_\mathcal{W}( \nu,r_\alpha (\nu))},
\end{equation}
then, for every $\tau \in {\cal T}_\alpha(\mu)$, we have
\begin{equation} \label{Eq.2v.2}
 \int \Wdd (P, \nu) \tau^* (P) \mu (d P ) 
\leq  
 \int \Wdd (P, \nu) \tau (P) \mu (d P ) .
\end{equation}
\end{Prop}
\noindent
{\bf Proof:} Let  $\tau \in {\cal T}_\alpha(\mu)$ and consider the real r.v. $X(P): = \Wdd (P, \nu)$. It is clear that the distribution of $X$, when we consider  in \Pd  the probability $\mu$ trimmed through the trimming function ${\tau^*}$, is stochastically  smaller than that associated to any other $\tau$. Therefore (\ref{Eq.2v.2}) holds.
\FIN

Note that equality in (\ref{Eq.2v.2}) is only possible if (\ref{Eq.2v.1}) happens for $\tau$. Thus, the optimal trimming functions must satisfy (\ref{Eq.2v.1})  where $\nu$ must be a barycenter of the trimmed probability associated to $\tau^*$. In other words, the optimal trimming functions are essentially defined by  the indicator of a ball centered at a trimmed barycenter.

We turn now to consistency of trimmed barycenters.  Theorem \ref{consistencytypeLoubes} (see subsection \ref{Sec.Consistency} for a proof) 
guarantees it under weak consistency of the probability distributions. Note that, unlike in the case of (non trimmed) barycenters, it is
not necessary that $\mathcal{W}_2(\mu_n,\mu)\to 0$, but it suffices to assume that $\mu_n\to_w \mu$.
 
 \begin{Theo}\label{consistencytypeLoubes}
Let $(\mu_n)_n, \mu $ be  probabilities { on $\Pd$} such that  $\mu_n \convw\mu$. 
For a fixed $\alpha \in(0,1)$, let $\bar \mu_n^\alpha$ be any trimmed barycenter of $\mu_n$. 
Then  the trimmed variances  converge, namely, $\mbox{\em Var}_\alpha(\mu_n)\conv \mbox{\em Var}_\alpha(\mu),$  
the sequence  $(\bar \mu_n^\alpha)_n$ is precompact  for the $\Wd$ topology and any 
limit is a trimmed barycenter of $\mu$. { If}  $\mu$ has only one 
trimmed barycenter, $\bar \mu^\alpha$, then  $\Wd(\bar \mu_n^\alpha,\bar \mu^\alpha)\to 0$.
\end{Theo} 

Repeating the argument that we use for law of large numbers for barycenters (Theorem \ref{consistency}), 
we obtain from Theorem \ref{consistencytypeLoubes} the corresponding one for trimmed barycenters. 
We state the result under the additional hypothesis of uniqueness of the trimmed barycenter of the probability law.  
{This kind of assumption  is quite common when showing consistency of centralization measures to avoid complicated  or too simplistic statements with complicated proofs even on $\Rea^k$. If the $\mu$-probability of the set of absolutely continuous probabilities in  $\Pd$ is greater than $1-\alpha$, the support of every $\alpha$-trimmed version of $\mu$ would contain absolutely continuous probabilities, thus it would have only one barycenter. Therefore, lack of uniqueness of the trimmed barycenter should be provoked by particular configurations of $\mu$. For example, for the uniform distribution on $[0,1]$, every point in the set $[(1-\alpha)/2,(1+\alpha)/2]$ is an $\alpha$-trimmed mean. Section 5 in Garc\'{\i}a-Escudero et al. \cite{CLT} treats this problem, although in practice it is quite rare to find distributions where uniqueness fails and, even then, the lack of uniqueness
could be only due to an improper choice of $\alpha$.}

\begin{Theo}\label{consistency2}
Assume that $\mu$ is a probability on the space \Pd with a unique trimmed barycenter. If $\mu_n$ is the sample probability giving mass $1/n$ to the probabilities $P_1,\ldots,P_n$ obtained as independent realizations of $\mu$, then the trimmed barycenters and variances are strongly consistent: $\bar \mu_n^\alpha \convs \bar \mu^\alpha,$ and $\mbox{\em Var}_\alpha(\mu_n) \convs \mbox{\em Var}_\alpha(\mu).$
\end{Theo}

\subsection{Location-scatter families}

Computation of Wasserstein distances and of barycenters for probabilities on the real 
line can be done through the explicit characterizations given in (\ref{casoreal}) and 
Proposition \ref{casoreal2}. In the multivariate setting, Proposition \ref{casoreal2}
can be extended to probabilities that can be parameterized 
in terms of a location and a scatter matrix, generalizing the normal multivariate model. 

\begin{Defi}\label{loc. scatter}
Let $ {\cal M}_{d\times d}^+$ be the set of $d\times d$ positive definite matrices   
and let $\bf{X}_0$ be a random vector with probability law  $P_0 \in \mathcal{P}_{2,ac}(\Read)$. 
The set $$\mathcal{F}(P_0):=\{\mathcal{L}(A{\bf X}_0+m): A\in {\cal M}_{d\times d}^+, m\in\Read\}$$  
of probability laws induced by  positive definite affine transformations from $P_0$ will be called a {\it location-scatter family}. 
\end{Defi}

As an easy consequence of Theorem \ref{Gelbrich2}, any probability $P \in \mathcal{F}(P_0)$ can be 
optimally transported to any other $Q \in \mathcal{F}(P_0)$ through an affine transformation with 
positive definite matrix. Thus w.l.o.g. we can assume that the mean of $P_0$ is the vector $\bar 0$ 
and its covariance matrix is $I_d$, the identity matrix. Also note that to make reference to a 
probability in $\mathcal{F}(P_0)$ we could use its mean $m$, and the transformation $A$ or alternatively 
$\Sigma=A^2$, the corresponding covariance matrix. We will use the second option to share the usual notation in the normal model. Therefore,  $\mathbb{P}_{m,\Sigma}$  will denote the probability in $\mathcal{F}(P_0)$ with mean $m$ and covariance matrix $\Sigma$.

In the statistical literature, a location and scatter family usually refers to an 
elliptical model. However, the families considered in this work under this denomination
include the elliptical families, but also families induced by different shapes. 
For instance, if we take in $\Rea^2$ the probability $P_0$ whose marginals are independent 
standard normal and exponential, respectively, then the family $\mathcal{F}(P_0)$ is not elliptical.
We also note that to address a confidence set problem, $P_0$  and the choice of any measurable set 
$\mathcal M_\gamma$ in \Read, such that $P_0(\mathcal M_\gamma)=\gamma$, will play the role of 
shape of the reference set. A typical asymptotic pivotal function for a  parameter $\theta \in \Read$ 
has the structure $n^{1/2}\hat{V}_n^{-1/2}(\hat{\theta}_n-\theta)$, thus, if we approximately know its law,   $P_0$,    then the set $\{\hat{\theta}_n-n^{-1/2}\hat{V}_n^{1/2}x: x\in \mathcal M_\gamma\}$ would be an approximate confidence set of level $\gamma$. Therefore the estimation of the location and scatter in the family $\mathcal{F}(P_0)$  produces a confidence set of the desired level, and a consensus based estimation would automatically produce a consensus confidence set for the parameter.

We show in Theorems \ref{extensionGelbrich} and \ref{self-suf} below
that Wasserstein barycenters and trimmed barycenters of probabilities supported on a location-scatter
family belong to the location-scatter family,
or, in other words, that location-scatter families {\it are closed for barycenters}. Of course, the general
equivariance result for similarity transformations (recall Proposition \ref{equiv}) remains true in 
the location-scatter setup. We also include
a  Gelbrich's type result showing that the dispersion in the \Wd-sense is minimized just when the probabilities belong to 
a common location-scatter family, in particular when all the probabilities are normal. The proof can be found in subsection \ref{locscale}.

\begin{Theo}\label{extensionGelbrich}
Let $\{P_i\}_{i=1}^k$ be probabilities in $\mathcal{P}_{2,ac}(\Read)$ with means $m_i, i=1,\ldots,k,$ and nonsingular covariance matrices $\Sigma_i, i=1,\ldots,k$. Let $N_i=N(m_i,\Sigma_i), i=1,\ldots,k,$ be normal probability distributions on $\Rea^d$. Also let $P_0 \in \mathcal{P}_{2,ac}(\Read)$ and let us denote by $\mathbb{P}_{m,\Sigma}$ the probability in $\mathcal{F}(P_0)$ with mean $m$ and covariance matrix $\Sigma$. 

Let us consider $\lambda_1,\ldots,\lambda_k$ positive weights with $\sum_{i=1}^k\lambda_i=1$, and respectively denote by $\bar P$, $\bar N$ and $\bar {\mathbb{P}}$ the (unique) barycenters of $\{P_i\}_{i=1}^k$, $\{N_i\}_{i=1}^k$ and $\{\mathbb{P}_{m_i,\Sigma_i}\}_{i=1}^k$. Then we have:
\begin{equation}\label{desigualdades}
\sum_{i=1}^k\lambda_i \Wdd(P_i,\bar P)\geq \sum_{i=1}^k\lambda_i \Wdd(\mathbb{P}_{m_i,\Sigma_i},\bar {\mathbb{P}})=\sum_{i=1}^k\lambda_i \Wdd(N_i,\bar N).
\end{equation}

 Moreover the inequality in (\ref{desigualdades}) can be an equality only if the mean and covariance matrix of $\bar P$ coincide with those of $\bar N$ and the relation $\{P_i\}_{i=1}^k \subset \mathcal{F}({\bar P})$ holds.
\end{Theo}

\begin{Nota}\label{remark1}
{\rm
We stress the fact that Theorem \ref{extensionGelbrich} generalizes (with the same proof but adding some notational complexity) to any $\mu \in W_2(\Pd)$ if, using the notation employed in (\ref{notacionbar}), we assume that for every $\omega\in\Omega,$   $\mu_\omega \in \mathcal{P}_{2,ac}(\Read)$  with mean $m_\omega$ and covariance matrix $\Sigma_\omega \in \mathcal{M}_{d\times d}^+.$
}
\end{Nota}

 \begin{Theo}\label{self-suf}
 Let  $P_0 \in \mathcal{P}_{2,ac}(\Read)$,  and $\mu \in  W_2(\Pd)$. With the notation in Remark \ref{remark1}, assume that  for every $\omega\in\Omega,$ the probability $\mu_\omega \in \mathcal{F}(P_0)$. Then the unique barycenter, $\bar \mu$, of $\mu$ also belongs to $\mathcal{F}(P_0)$. The mean of $\bar \mu$ is $\bar m:=\int m_\omega \Prob(d\omega),$ and the covariance matrix, $\bar \Sigma$, is the only positive definite matrix satisfying 
 \[
\bar \Sigma=\int\left(\bar \Sigma^{1/2}\Sigma_\omega\bar \Sigma^{1/2}\right)^{1/2}\Prob(d\omega)
\]
 \end{Theo}
\medskip

Once we know that a family is closed for barycenters,  the property will be shared by the trimmed barycenters. This is motivated by the fact that trimmed versions of a probability $\mu$ have their supports contained in that of $\mu$, and a trimmed barycenter is characterized as a barycenter of an optimal trimmed version of $\mu$.  Once a trimming function has been fixed, the uniqueness of the barycenter of absolutely continuous distributions, obtained in \cite{Carlier}, leads also to the uniqueness of the trimmed barycenter associated to that trimmed version of $\mu.$   However, we cannot deduce  uniqueness of the trimmed barycenters in an easy way. In fact this is a hard problem even for  trimmed means in  euclidean spaces.

\begin{Coro}\label{self-suf2}
{ Assume that  $P_0 \in \mathcal{P}_{2,ac}$  and $\mu$ a probability on $\Pd$ that is supported in $\mathcal{F}(P_0)$.} Then, for every $\alpha \in (0,1)$, any trimmed barycenter of $\mu$ also belongs to $\mathcal{F}(P_0).$ Moreover any optimal trimming function for $\mu$ uniquely determines a trimmed barycenter.
 \end{Coro}
 
 \begin{Nota}\label{remark2}
 {\rm
  We emphasize the importance of  Corollary \ref{self-suf2} that allows to search for trimmed barycenters of, say a random normal distribution, looking just to the means and covariance functions.  Moreover, by Theorem \ref{Gelbrich2}, the distance between probabilities in $\mathcal{F}(P_0)$ is given by
\begin{equation}\label{distancelocscale}
\Wdd(\mathbb{P}_{m_1,\Sigma_1},\mathbb{P}_{m_2,\Sigma_2})=
\|m_1-m_2\|^2+\mbox{trace}\Big(\Sigma_1+\Sigma_2-2\Big(\Sigma_1^{1/2}\Sigma_2\Sigma_1^{1/2}\Big)^{1/2}\Big) ,
\end{equation}
which allows computation of Wasserstein distances. With applications in view, these facts will be complemented with the proposal of a feasible algorithm for addressing the computation of the trimmed barycenter of a finite set of probabilities that belong to a location-scatter family and a given set of weights.
  }
 \end{Nota}
 
\begin{figure}[tb] 
\begin{center}
\includegraphics[width=7.5cm]{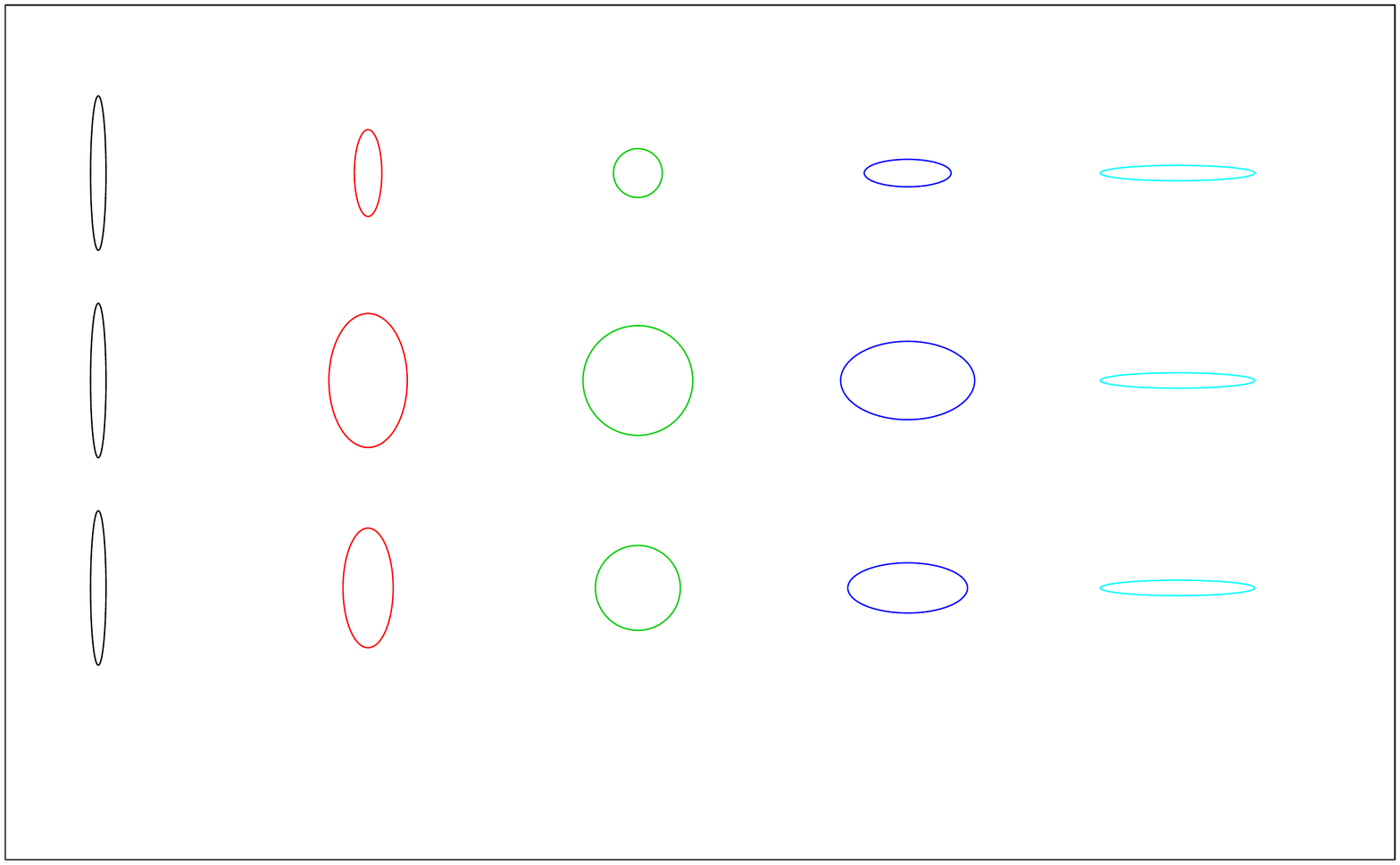}
\includegraphics[width=7.5cm]{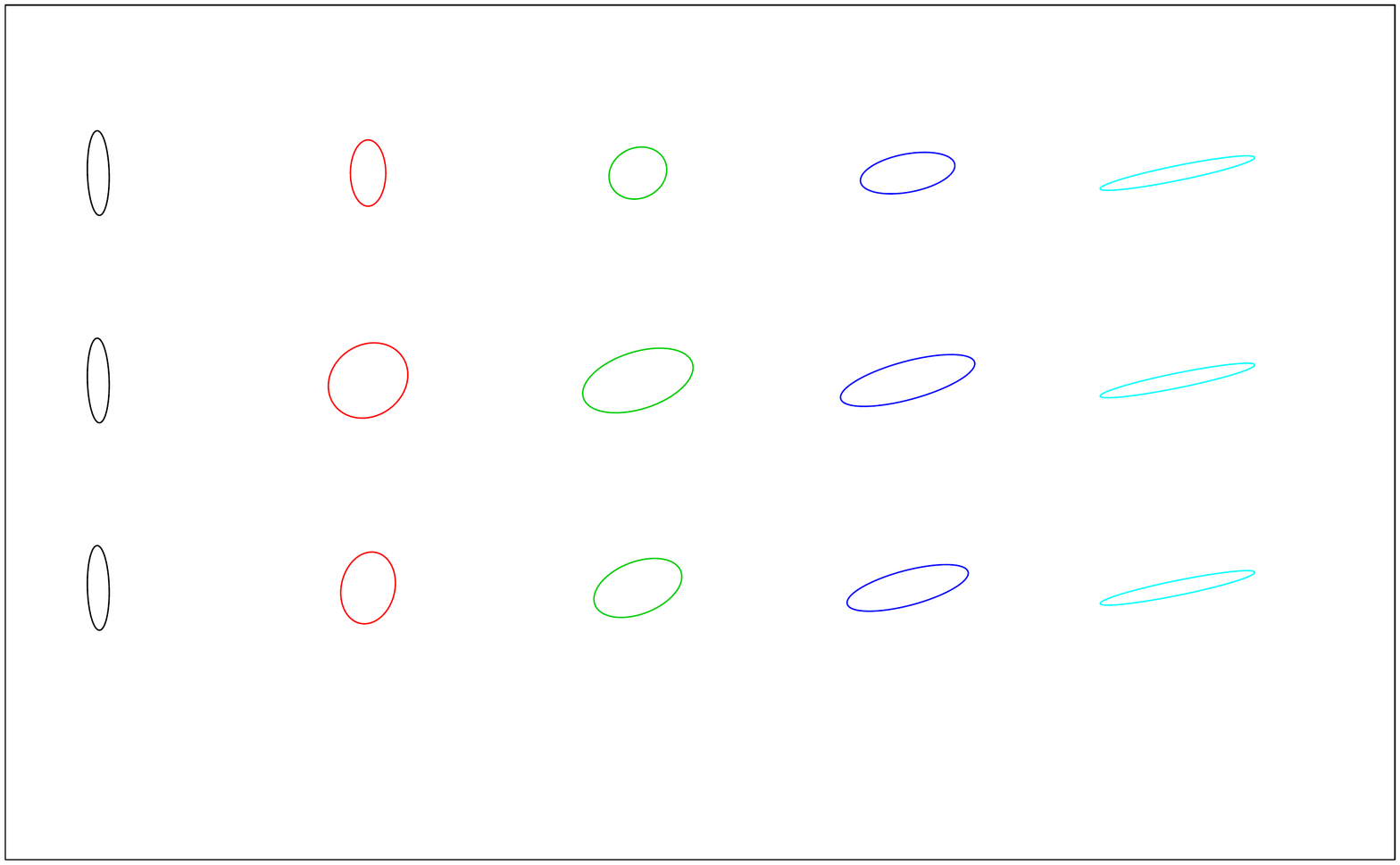}
\vspace{-1cm}
\caption{Each picture shows the effects of linear interpolation of two matrices corresponding to the weighted mean (middle row), barycenter (lower row) and Log-Euclidean mean (upper row), of the matrices represented  by  the black and cyan ellipses. From left to right we handle the weights  0, .75, .50, .25, 1 on the black one. Note the characteristic swelling effect associated to the weighted mean.}
\label{BarvsLog}
\end{center}
\end{figure}

\begin{figure}[tb] 
\begin{center}
\includegraphics[width=7.5cm]{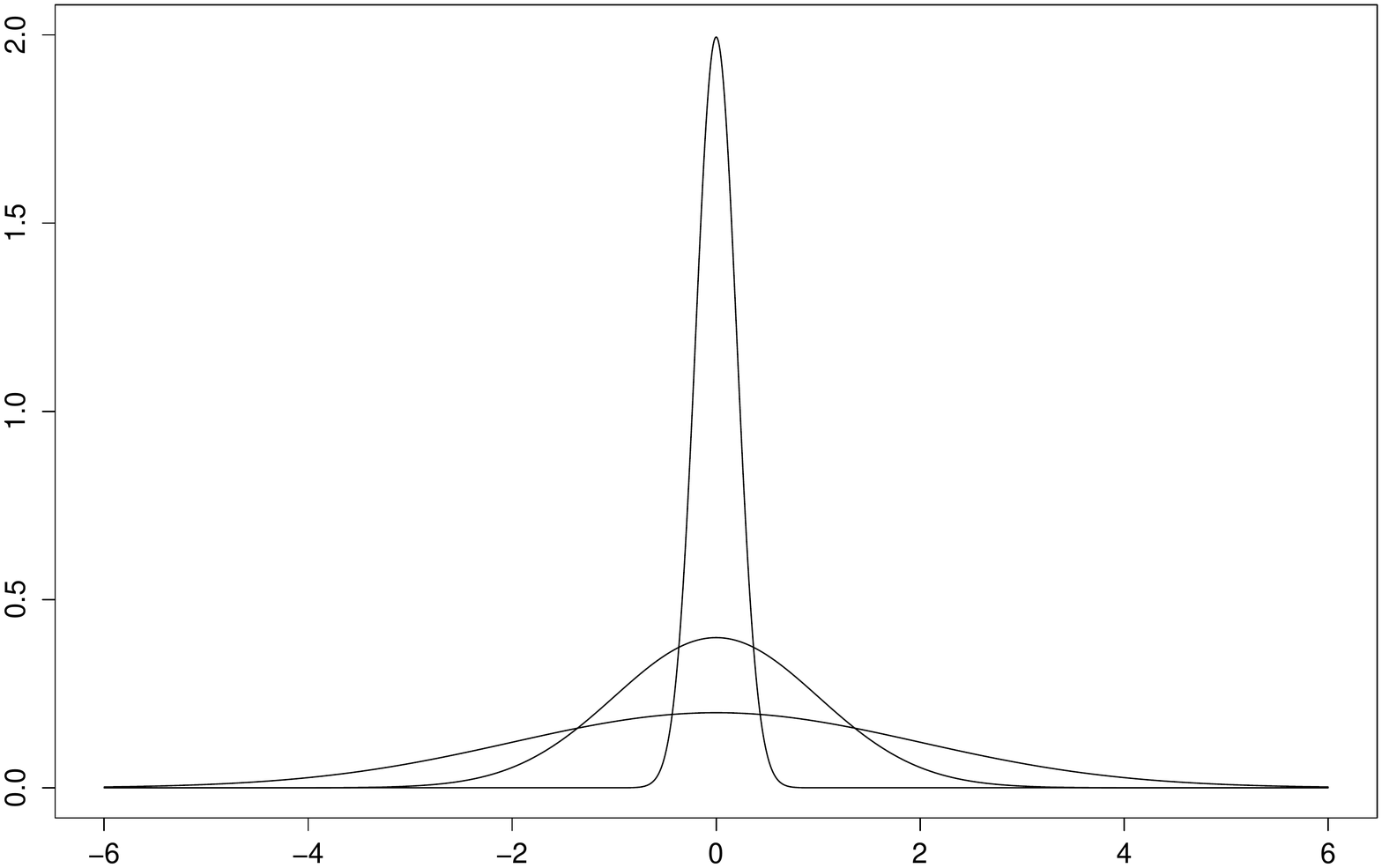}
\includegraphics[width=7.5cm]{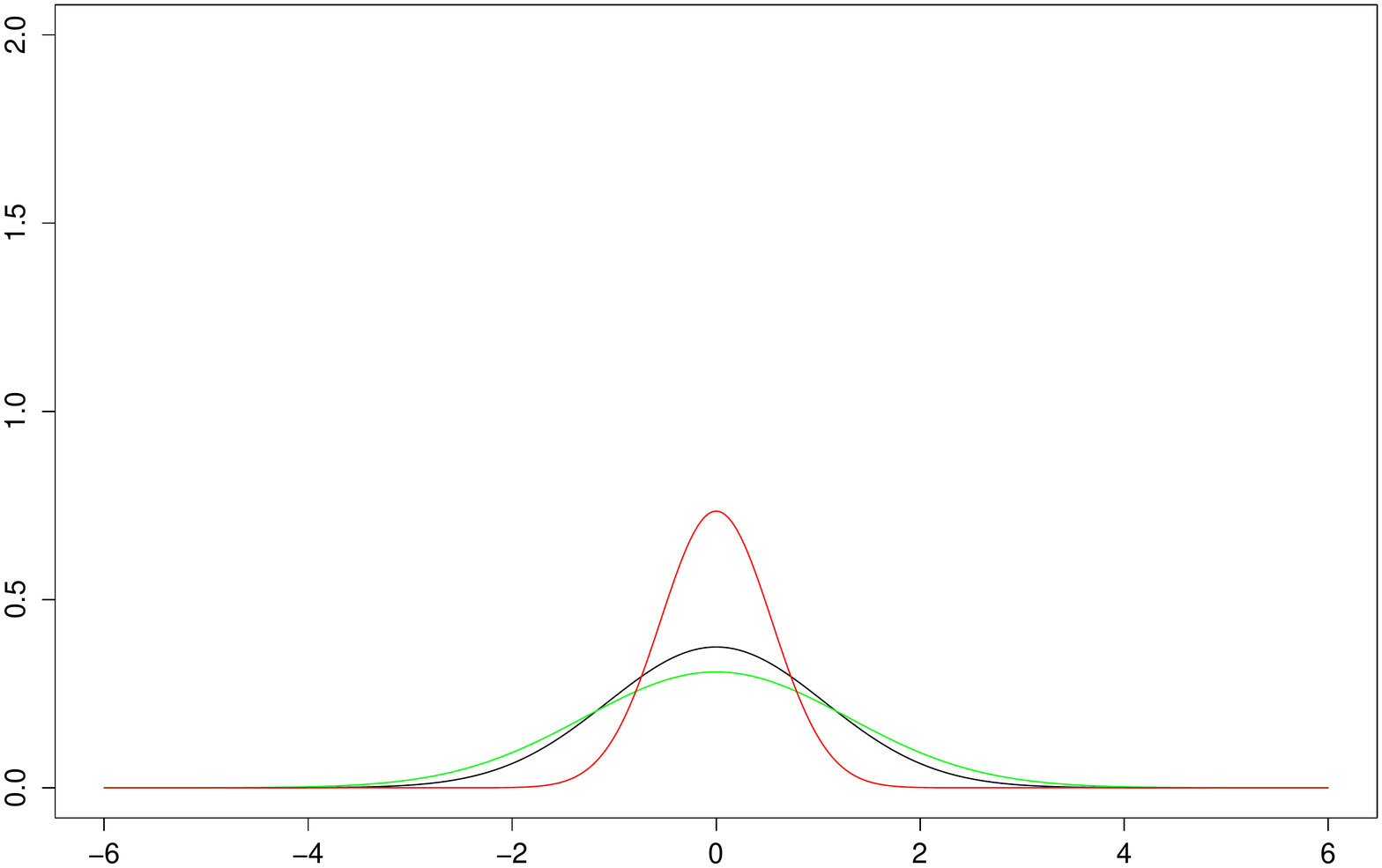}
\vspace{-.5cm}
\caption{Left: Density functions corresponding to $N(0,\sigma_i)$ distributions for $\sigma_1=.2, \sigma_2=1, \sigma_3=2$. Right: Normal density functions for $\sigma=\left(1/3\sum_{i=1}^3 \sigma_i^2\right)^{1/2}=1.296$ (green), $\sigma= \left(\prod_{i=1}^3 \sigma_i\right)^{1/3}=.737$ (red) and $\sigma=1/3\sum_{i=1}^3 \sigma_i=1.067$ (black), respectively associated to the mean of variances and the geometric mean (or Log-Euclidean)  of the variances, and the barycenter of the distributions.}
\label{BarvsLoginR}
\end{center}
\end{figure}

Once this theory has been  developed it can be argued that  (\ref{distancelocscale}) is just a combination of metrics: the Euclidean metric
for the means plus another one between covariance matrices. Since the final product only involves distributions in $\mathcal{F}(P_0)$, even some comparison with combinations of other metrics should be in order. Focusing on the metric on the covariance matrices,  Fr\'echet means related to several metrics on this space of symmetric positive definite matrices have been proposed in the literature. Among these metrics particular attention is deserved by the affine-invariant metrics and Log-Euclidean metrics, introduced by considerations that mainly arise from the image analysis framework (see Arsigny et al. \cite{LogEuc}). In both cases, the associated Fréchet means can be considered as generalizations of the geo\-metric mean, although the Log-Euclidean mean could be preferred by its easier computation.
 We should note that our choice of (\ref{distancelocscale}) is not guided by the search for a metric on this set of matrices, but it is rather the restriction of a metric on the set of all probabilities
with finite second moment –a kind of $L_2$ space– with suitable properties already pointed out in the literature in different scenarios. We note also that the computation of Wasserstein barycenters can be efficiently done through the algorithm introduced in \cite{preprint}  and discussed in Section \ref{computation}. Taking this  into account, the comparison must rely on purely statistical arguments, like those involving the comparison between the mean and the geometric mean for real numbers. Any of them can be preferred for different tasks but, arguably, the usual mean is the preferred choice in most of the applications. To provide some illustrative  idea of their relative behavior, in Figure \ref{BarvsLog} we resort to the comparison of the interpolation of two pairs of covariance matrices represented by the black and cyan ellipses in each picture. Notice that the average (the weighted mean of covariances) is included for  reference. The upper, middle and lower rows are respectively associated to Log-Euclidean, average and barycenter approaches. The red, green and blue ellipses respectively represent the solutions associated to .75, .5 and .25 weights on the black covariance matrix (and .25, .5 and .75 on the cyan one). Additionally, we include in Figure \ref{BarvsLoginR} the  density functions of three centered normal distributions accompanied by those associated to these approaches.
For very similar standard deviations $\{\sigma_i\}_{i=1}^k$ and any associated weights  $\{\lambda_j\}_{ j=1}^k,$ the three aggregation procedures would produce nearly the same result but, if this is not the case, the estimates can be very different. 

An explanation for these different behaviors comes from Jensen's inequality.  In the simplest one-dimensional case, these three averaging procedures result in standard deviations given by  the left (Log-Euclidean), middle (Wasserstein barycenter) and right (weighted average of variances) terms in the following inequalities
\begin{equation}\label{desigualdades}
\mbox{exp}\left(\sum_{j=1}^k\lambda_j\log \sigma_j\right)\leq \sum_{j=1}^k\lambda_j\sigma_j\leq\sqrt{\sum_{j=1}^k\lambda_j\sigma_j^2}.
\end{equation}
This shows that the standard deviation of the geometric mean is smaller than the average of the standard deviations which in turn is smaller than the standard deviation arising from the weighted mean of the variances. This gives some explanation to the swelling effect associated to the weighted mean. We also note that if we are willing to admit that the standard deviation $(\int |x|^2P(dx))^{1/2}$  is a good measurement of the size of a centered distribution, $P$, then the Log-Euclidean mean results in summaries which are smaller than the average size of the objects to be summarized. In this sense, the Wasserstein barycenter provides the better choice between these alternatives.

For diagonal (in some basis) covariance matrices, this explains the intermediate size of the barycenter, avoiding the 
swelling effect of the mean of variances, but also the somewhat excessive decrease associated to the Log-Euclidean approach. In a location scatter model, for a finite collection $\{\mathbb{P}_{0,\Sigma_j}\}_{j=1}^k$ and weights $\{\lambda_j\}_{j=1}^k,$ and the principal directions of the $\Sigma_j$ matrices are the same, then for some orthonormal matrix $H$, $\Sigma_j=HD_jH^t, j=1,\dots,k$ with $D_j=\mbox{diag}(\sigma^2_{j1},\dots,\sigma^2_{jd})$. If we denote by $\Sigma^*,\bar \Sigma, \widehat \Sigma$ the covariance matrices associated to the Log-Euclidean, Wasserstein barycenter and weighted average approaches, then also $\Sigma^*=HD^*H^t,\bar \Sigma=H\bar DH^t, \widehat \Sigma=H\widehat DH^t,$ with $D^*=\mbox{diag}({\sigma^*_1}^2,\dots,{\sigma^*_d}^2),$ $\bar D=\mbox{diag}({\bar \sigma_1}^2,\dots,{\bar \sigma_d}^2),$ $\widehat D=\mbox{diag}({\widehat \sigma_1}^2,\dots,{\widehat \sigma_d}^2),$ which are   related by $\sigma^*_j \leq \hat \sigma_j \leq \widehat \sigma_j, j=1,\dots,d,$ from (\ref{desigualdades}) because $\sigma^*_j=\exp \left(\sum_{i=1}^k\lambda_i\log \sigma_{ji}\right)$, $\hat \sigma_j=\sum_{i=1}^k\lambda_i\sigma_{ji}$ and ${\widehat{\sigma_j}}^2=\sum_{i=1}^k\lambda_i{\sigma_{ji}}^2$. Also note that in this case we obtain again that  the ``standard deviation" of the Barycenter is the weighted mean of the standard deviations. 

Although the fact just noticed will be not true in full generality, we will show below that such weighted mean of standard deviations is an upper bound for the standard deviation of the barycenter. We would like to stress that this result will be proved for probabilities that do not necessarily belong to a location-scatter family. Even more, by Remark 3.4 in \cite{preprint}, the property is true even without the absolutely continuous assumption that we will impose here for a simpler argument.

\begin{Prop}\label{standardDev}
Let $P_1,\dots,P_k\in \mathcal P_{2,ac}(\Read)$ centered in mean, and 
$\lambda_1,\dots,\lambda_k$ be positive weights adding one. If $\bar P$ is the associated barycenter, then $$\left(\int \|x\|^2\bar P(dx)\right)^{1/2} \leq \sum_{j=1}^k\lambda_j \left(\int \|x\|^2 P_j(dx)\right)^{1/2}.$$
\end{Prop}

\section{Computation of Barycenters and Trimmed Barycenters}\label{computation}
 
The characterization of trimmed barycenters given in Proposition \ref{Lemm.charac2} leads to consider the effective computation of barycenters as a first step in the obtention of trimmed barycenters. We recall  for probabilities on $\Rea$ the characterization given in Proposition \ref{casoreal2} in terms of quantiles. If $\mu$ is the probability on $\mathcal{P}_2(\Rea)$ giving weights $\lambda_1,\dots,\lambda_k$ to the probabilities $P_1,\dots,P_k$, then the barycenter $\bar\mu$ is the distribution of the random variable $\sum_{j=1}^k\lambda_jF_j^{-1}$  (defined on the unit interval), thus denoting by $m_j$ and $\sigma_j^2$ the mean and variance of $P_j$, and $\bar m$ and $\mbox{Var}(\bar\mu)$ those of $\bar \mu$:
\begin{equation}\label{varianzauniv}
\mbox{Var}(\mu)= \sum_{j=1}^k\lambda_j(m_j-\bar m)^2+\sum_{j=1}^k\lambda_j\sigma_j^2-\mbox{Var}(\bar\mu).
\end{equation}
 When $P_1,\dots,P_k$ belong to a common location-scale family, $\mathcal{F}(P_0)$, where $P_0$ has quantile function  $F_0^{-1}$ (with zero mean and variance 1), then $F_j^{-1}=m_j+\sigma_jF_0^{-1},\ j=1,\dots,k,$ and with $\bar\sigma:=\sum_{j=1}^k\lambda_j\sigma_j$, (\ref{varianzauniv}) specializes to
\[
\mbox{Var}(\mu)= \sum_{j=1}^k\lambda_j(m_j-\bar m)^2+\sum_{j=1}^k\lambda_j(\sigma_j-\bar\sigma)^2=\sum_{j=1}^k\lambda_j(m_j^2+\sigma_j^2)-(\bar m^2+\bar \sigma^2).
\]

In contrast, as previously noted, in the multivariate case closed expressions are only available  just for situations 
essentially equivalent to several univariate cases. This  is the case if, e.g. the probabilities share a common structure 
of dependence in some particular basis (see Section 2 in Cuesta-Albertos et al \cite{Cuesta-Albertos1993} or Section 4 in \cite{Bois15}), or if they are 
radial transformations of a common probability law (see Section 3 in \cite{Cuesta-Albertos1993}). 
Turning to approximate computations, in recent times some papers addressed the goal of numerical computation of Wasserstein barycenters, see Cuturi and Doucet \cite{Cuturi}, Benamou et al. \cite{Benamou} or Carlier et al. \cite{Carlier}. In these cases, the approaches address the case of sample distributions or are based on the discretization of the problem through a fine grid and the use of suitable optimization procedures. Although their results allow to get good representations for the barycenter of distributions with very different shapes, the grid sizes for suitably approximating the distributions must be large and would strongly depend on the dimension making them highly time-consuming even in small dimensions and with a small number of distributions. Of course these procedures allow computation of barycenters, but regrettably, under trimming, the available methods to compute the trimmed barycenters (even for real random variables), like our Algorithm for the trimmed barycenter below, need several initializations and often require the iterative computation of several thousands of barycenters.  This makes  those algorithms based on discretizations to be, by now, inapplicable  for our proposes. Fortunately, for one of the most important cases in multivariate statistics, namely the location-scatter families,
a fast  consistent procedure for approximating the numerical solution of equation (\ref{ecuacion}) has recently been introduced in \cite{preprint}.
We give here a quick description of the procedure.

Assume that $
P_1,\ldots,P_k\in\mathcal{P}_{2,ac}$
 and the weights $\lambda_1,\ldots,\lambda_k$ are fixed. Given $\eta \in \Pd$, we consider the functional
$$V(\eta):= \sum_{i=1}^k \lambda_i \mathcal{W}_2^2 (\eta,P_j),$$
looking for $\bar P \in \Pd$ such that
$$V(\bar{P})=\min_{\eta\in \subPd}V(\eta).$$

If $\eta\in\mathcal{P}_{2,ac}$, we know that there exist optimal transport maps $T_j$ from $\eta$ to $P_j$. Assume that $X$ is a random vector with law $\eta$, thus $$\mathcal L(T_j(X))=P_j, \mbox{ and } \mathcal W_2^2(\eta,P_j)=\Exp\|X-T_j(X)\|^2, \ j=1,\dots,k.$$

With this notation we define
\[
G(\eta):=\mathcal{L}\Big(\sum_{j=1}^k \lambda_j T_j(X)\Big),
\]
 to design a consistent, iterative procedure for the approximate computation of $\bar{P}$. 
Next, we collect some basic properties of $G$ that show a link between the $G$ transform and the barycenter problem.

\begin{Prop}\label{basicineq}
If $\eta\in\mathcal{P}_{2,ac}$ then
\[
V(\eta)\geq V(G(\eta))+\mathcal{W}_2^2(\eta,G(\eta)).
\]
In particular,   if the barycenter, $\bar P$, is absolutely continuous then $G(\bar P)=\bar P$.
\end{Prop}

We remark that the hypothesis of absolute continuity of $\bar P$ is required just to guarantee that $G(\bar P)$ is defined. The theory developed for the location-scatter families, and particularly for normal distributions, allows to guarantee this in such cases. On the other hand, the conclusion of the proposition invites to consider an iterative process, starting from any $\eta_0\in \mathcal{P}_{2,ac}$ and considering the sequence
\begin{equation}\label{iteration}
\eta_{n+1}:=G(\eta_n),\quad n\geq 0.
\end{equation}
We have proved the consistency of this iterative procedure  in greater generality in \cite{preprint}, but for our present purposes it suffices that given in the following statement.

\begin{Theo}\label{Gaussiancase}
If $P_1,\ldots,P_k$ are nonsingular Gaussian distributions on $\mathbb{R}^d$
and the initial measure, $\eta_0$, is also a nonsingular Gaussian distribution, then the iteration
defined by (\ref{iteration}) is consistent, namely,
$$\mathcal{W}_2(\eta_n,\bar{P})\to 0,$$
as $n\to\infty$, where $\bar{P}$ is the (unique) barycenter of $P_1,\ldots,P_k$.
\end{Theo}

It is time to recall  Theorem \ref{self-suf} on  barycenters of location-scatter families. We know from it that, given positive definite 
matrices $\Sigma_1,\ldots,\Sigma_k$ there exists a unique
positive definite matrix $\bar{\Sigma}$ { solving (\ref{ecuacion}).}

Reading Theorem \ref{Gaussiancase} just in terms of approximating the unique solution
of (\ref{ecuacion}), the conclusion becomes that if, starting from any  positive definite matrix
$S_0$, according to Theorem \ref{Gelbrich2}, we define
\begin{equation}\label{recur}
S_{n+1}=S_{n}^{-1/2}\Big(\sum_{j=1}^k \lambda_j(S_{n}^{1/2}\Sigma_j S_{n}^{1/2})^{1/2} \Big)^2 S_{n}^{-1/2},
\end{equation}
then 
$$
\lim_{n\to\infty} S_n=\bar{\Sigma}.
$$

Therefore the process leads to  a consistent iterative method for approximating
the solution of (\ref{ecuacion}). The method is easily { implemented} and, 
in practice, shows a very good performance. We refer to \cite{preprint} for further details.

The characterization of the distance between probabilities in the location-scatter family (\ref{distancelocscale}) leads to identical distances to those between normal laws with same location and covariance matrices. Therefore we can extend Theorems \ref{casonormal}   and \ref{Gaussiancase} in the following way.

\begin{Theo}\label{resumen}
If $\mu$ is the probability on $\Pd$ giving weights $\lambda_1,\dots,\lambda_k$ to the probabilities $\mathbb{P}_{m_1,\Sigma_1},\dots,\mathbb{P}_{m_k,\Sigma_k}\in \mathcal{F}(P_0)$, a location-scatter family with $P_0\in \mathcal{P}_{2,ac}(\Read)$, then its barycenter  is the probability  $\mathbb{P}_{\bar m,\bar \Sigma} \in \mathcal{F}(P_0),$ where $\bar m=\sum_{i=1}^k\lambda_im_i$ and $\bar \Sigma$ is the only definite positive matrix satisfying  equation (\ref{ecuacion}). Moreover, $\bar \Sigma$ can be obtained as the limit of the sequence defined in (\ref{recur}). The variance of $\mu$ takes the value
\begin{eqnarray*}
\mbox{Var}(\mu)
&=&
\sum_{j=1}^k\lambda_j\|m_j-\bar m\|^2+\sum_{j=1}^k\lambda_j\mbox{trace}(\Sigma_j-\bar\Sigma)
\\ 
&=&
\sum_{j=1}^k\lambda_j(\|m_j\|^2+\mbox{trace}(\Sigma_j))-(\|\bar m\|^2+\mbox{trace}(\bar \Sigma)).
\end{eqnarray*}
\end{Theo}

Through Theorem \ref{resumen} we can compute barycenters and variances for any finite set of probabilities and weights, once we know the corresponding locations $m_1,\dots,m_k$ and covariance matrices $\Sigma_1,\dots,\Sigma_k$. Moreover, the distances between probabilities are also easily computed through (\ref{distancelocscale}), which is valid for every location-scatter family. Therefore, Corollary \ref{self-suf2} and the characterization of the best trimming functions given in Proposition \ref{Lemm.charac2} allow to search for a trimmed barycenter  as the barycenter based on subsets of $P_1,\dots,P_k$ with an accumulate weight of at least $1-\alpha$ and minimum variance after normalizing the weights.


Next, we include an algorithm to obtain the trimmed barycenter of  the probabilities $\mathbb{P}_{m_1,\Sigma_1},\dots,\mathbb{P}_{m_k,\Sigma_k}\in \mathcal{F}(P_0)$ with weights $\lambda_1,\dots,\lambda_k$.  It combines estimation and concentration steps, being an adaptation of  usual algorithms for obtaining best (in some sense) trimmed regions, like the ones involved in the MCD or LTS robust estimators, with  the ne\-cessary updates of the distances and weights in each concentration step. Once an initial solution is provided, this kind of algorithm guarantees  convergence through the estimation and concentration steps, but we must also consider the possibility of local optimizers, a fact that leads to consider random choices of initial candidates to be compared at the end. We simply emphasize the fact that this algorithm shares the good performance of the versions currently used in similar problems on estimation in the multivariate setting.
 \medskip

\textbf{The algorithm}\label{algorithm}  
\begin{enumerate}
\item[0.] Fix $n=0$, and randomly choose initial candidates $\hat m_n, \hat \Sigma_n$ for the mean and the covariance matrix.
\item Compute the distances $d_i^n$ between $\mathbb{P}_{\hat m_n,\hat \Sigma_n}$ and $\mathbb{P}_{m_i,\Sigma_i}, i=1,\dots,k,$  through (\ref{distancelocscale}).
\item Consider the permutation $((1),\ldots,(k))$ such that  $d_{(1)}^n\leq \ldots \leq d_{(k)}^n$.
\item 
Set $j_n = \inf \{j:  \sum_{i\leq j} \lambda_{(i)} \geq 1- \alpha\}$ and define the new weights:
\[
\lambda_{(i)}^n=
\left\{
\begin{array}{ll}
\lambda_{(i)} & \mbox{ if } i < j_n
\\
1- \alpha - \sum_{ i < j_n} \lambda_{(i)}^n  & \mbox{ if } i = j_n
\\
0 &  \mbox{ if } i >  j_n.
\end{array}
\right.
\]

\item 
Since  $\sum_{i=1}^k \lambda_{(i)}^n  = 1 - \alpha$, define $\lambda_{(i)}^n=(1-\alpha)^{-1} \lambda_{(i)}^n,$ in order to have  $\sum_{i=1}^k \lambda_{(i)}^n  =1$.

\item Using the updated weights, compute $\hat m_{n+1}$, the weighted mean of the means, 
and $\hat \Sigma_{n+1}$  through the recursive algorithm (\ref{recur}).

\item { Iterate steps 1 through 5} until convergence.
\item Compute the variance of the final trimmed sample of probabilities and weights.
\item Go to  0 and finalize after a moderate number of initial choices, reporting the barycenter producing the minimum variance.
\end{enumerate}

As  a toy illustration of the results of the computation of the barycenters (trimmed or not), we present now two examples, in which we handle 2-dimensional normal distributions, allowing a suitable visualization of the results. In these examples, we represent graphically a normal distribution with mean $m$ and covariance matrix $\Sigma$ by the set
\[
\{ x \in \Rea^2 : (x-m)^t \Sigma^{-1} (x-m) =1\}.
\]

\begin{Ejem}{\rm
{ We have considered first} the six normal distributions represented in the graph in the left hand side in  Figure \ref{Elipses_1}.  We have computed the barycenter, and the $1/6$ and $2/6$ trimmed barycenters of these normal distributions. The results appear in the right hand side graphic. All three barycenters are normal distributions which are represented by the black, blue and red ellipses in the right hand side graphic in Figure \ref{Elipses_1}.

\begin{figure}[tb] 
\begin{center}
\includegraphics[width=7cm]{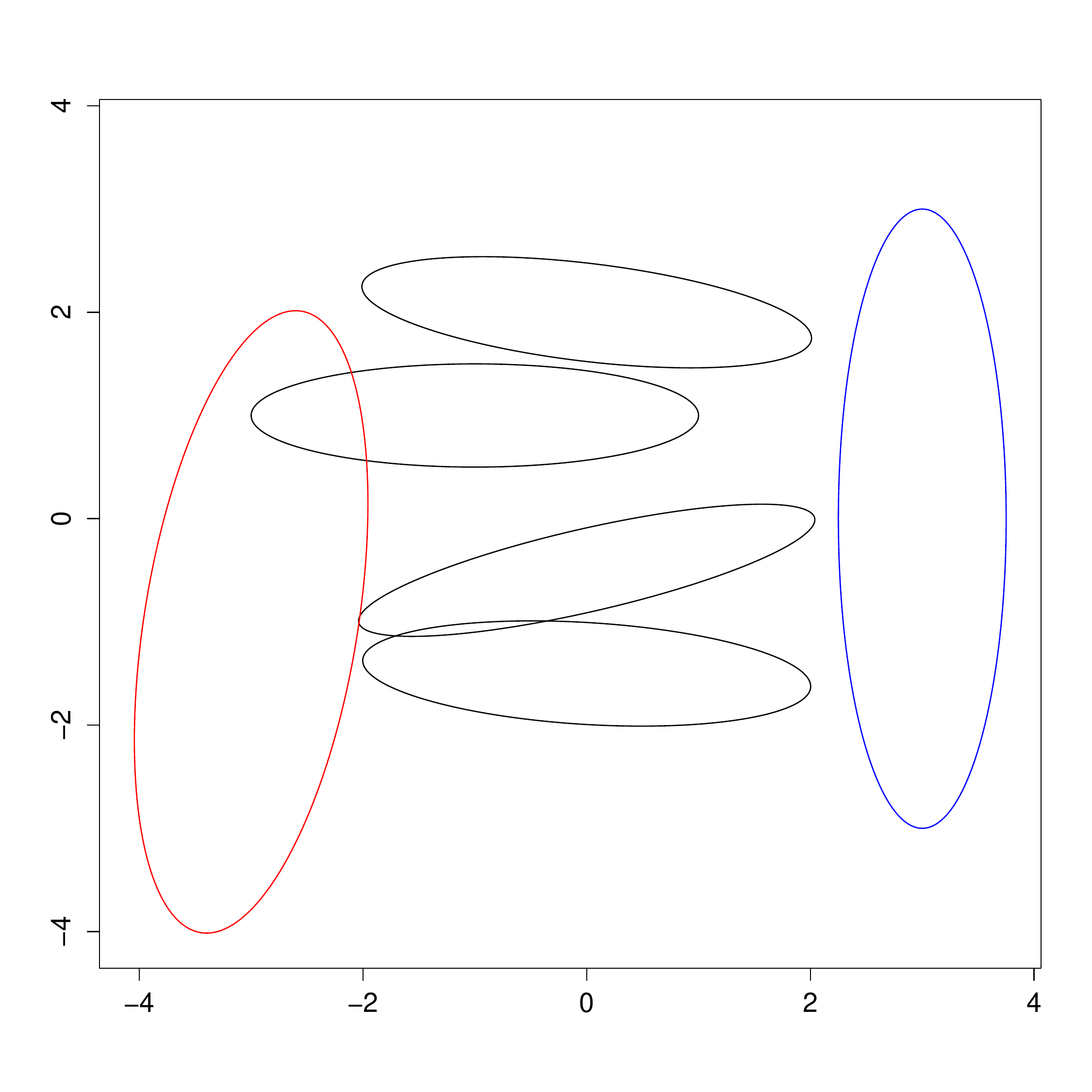}
\includegraphics[width=7cm]{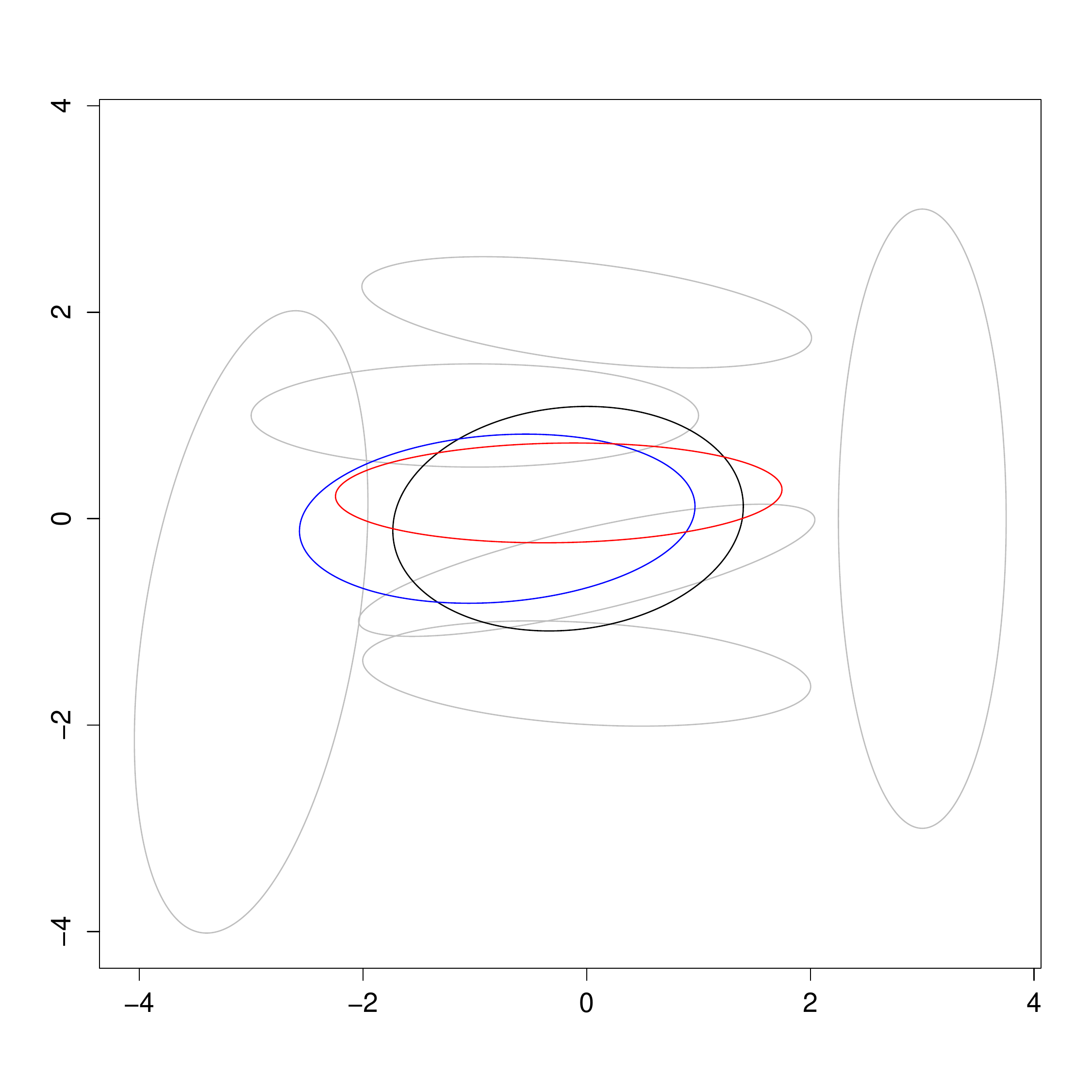}
\caption{Computation of trimmed barycenters (right figure) of the ellipses shown in the figure in the left}
\label{Elipses_1}
\end{center}
\end{figure}

The black ellipse is the non-trimmed barycenter.  Trimming  $\alpha = 1/6$ the barycenter is the blue ellipse, and the procedure trims the blue ellipse in the left graphic. 
The red ellipse shows the result of trimming $\alpha =2/6$. In this case, the procedure trims the red and the blue ellipses in the left hand side graphic. Observe that the red ellipse lies in  the middle of the four black ellipses in the left graphic showing a very similar  shape.

The previous result could  { have been anticipated}  because, according to (\ref{distancelocscale}), the decision of which distributions to trim depends on the shape and the location of the ellipses under consideration and, in this example,  the colored ellipses have different shapes and separated locations than the others. Because of this, we also show a not too big modification of this example which is shown in Figure \ref{Elipses_2}. Here five ellipses coincide with the corresponding ones in Figure \ref{Elipses_1}. However, the green ellipse in the left hand graph in this figure is one of the ``horizontal" ellipses whose center has been moved two units along the ordinates axis. Now, it happens that the trimmed distribution when taking $\alpha = 1/6$ continues being the blue one, but when taking $\alpha = 2/6$ the procedure trims the blue and the green ellipses leaving the red one untrimmed.

\begin{figure}[tb] 
\begin{center}
\includegraphics[width=7cm]{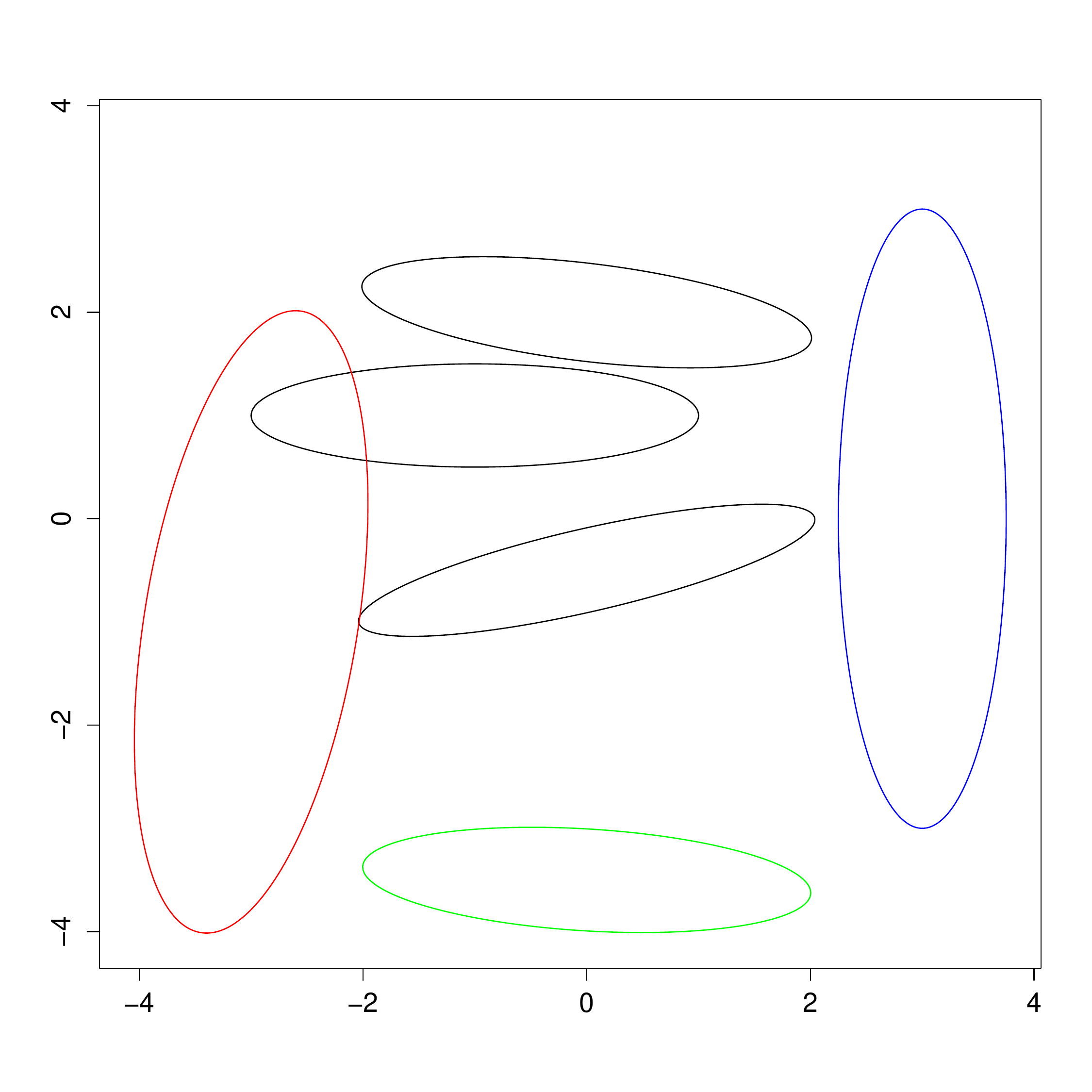}
\includegraphics[width=7cm]{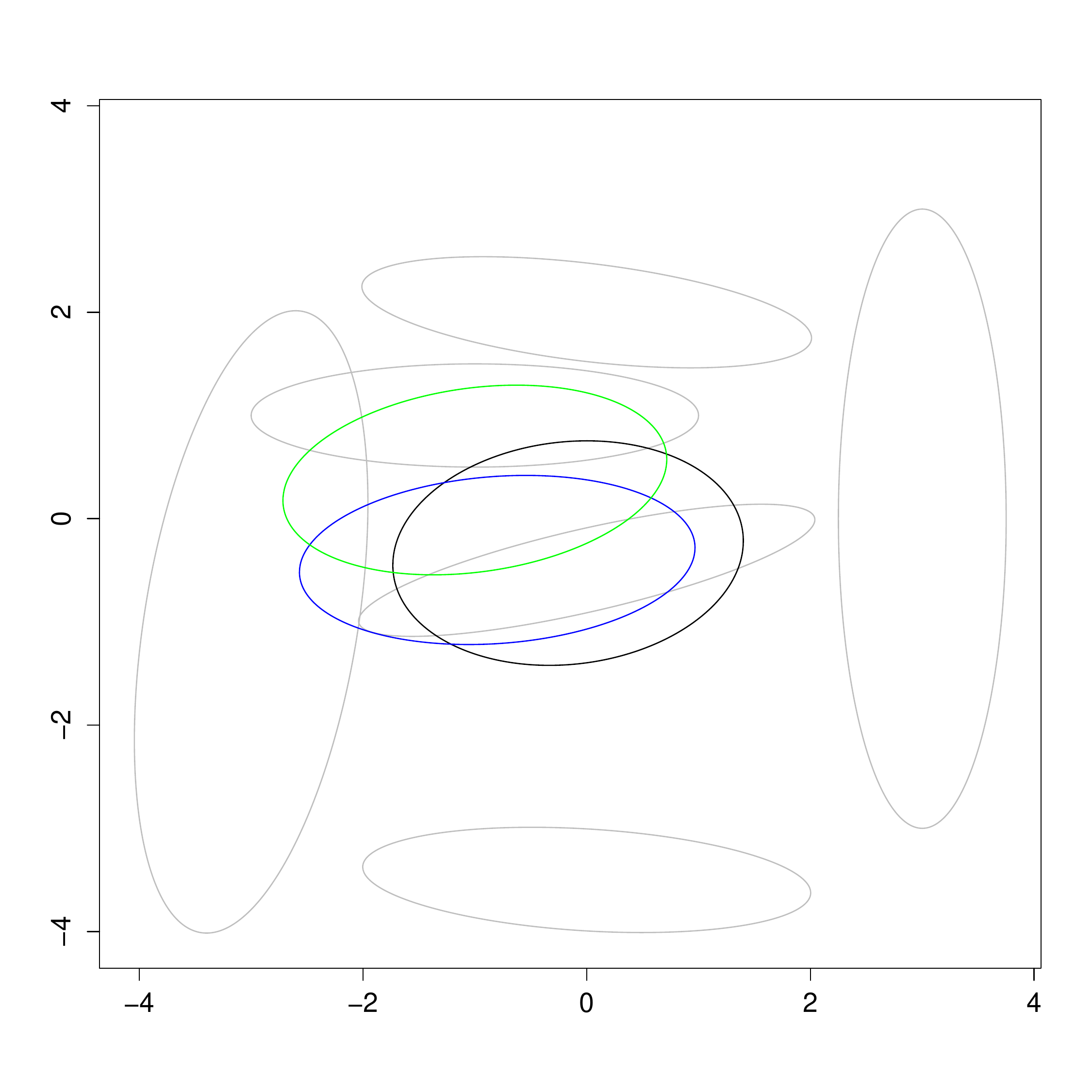}
\caption{Computation of trimmed barycenters (right figure) of the ellipses shown in the figure in the left}
\label{Elipses_2}
\end{center}
\end{figure}

}
\end{Ejem}

\begin{figure}[tb] 
\begin{center}
\includegraphics[width=12cm]{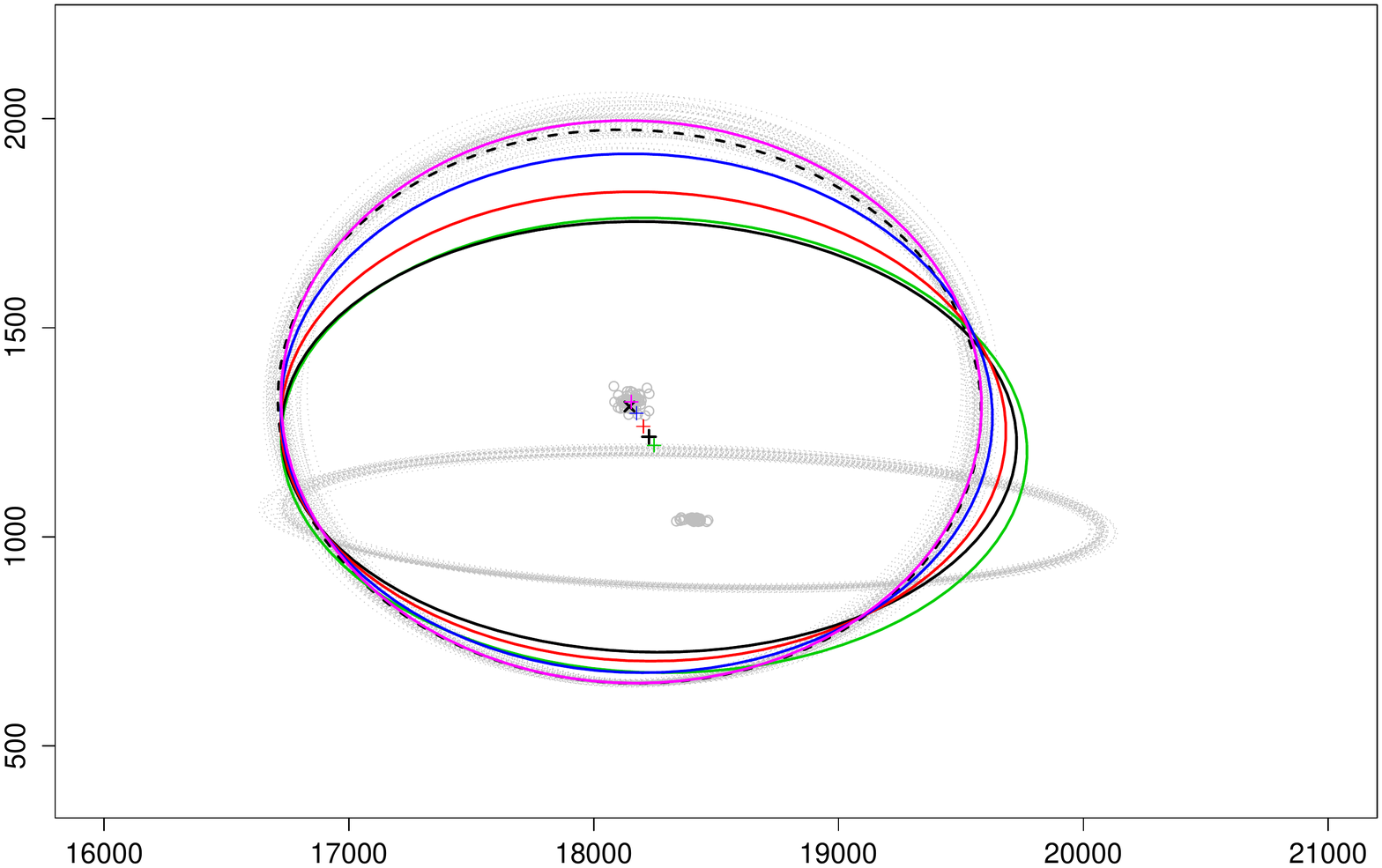}
\vspace{-.8cm}
\caption{Graph summarizing the MCD´s solutions obtained from 100 subsamples of size 5000 (gray) and those provided by several aggregations of the solutions and that given by the full Palomar Data set (black dashed). The green ellipse is associated to the mean of the solutions;  those in black, red, blue and magenta correspond to trimmed barycenters with respective trimming sizes .1, .2, .3 and .4.  }
\label{palomar}
\end{center}
\end{figure}

\begin{figure}[tb] 
\begin{center}
\includegraphics[width=7.5cm]{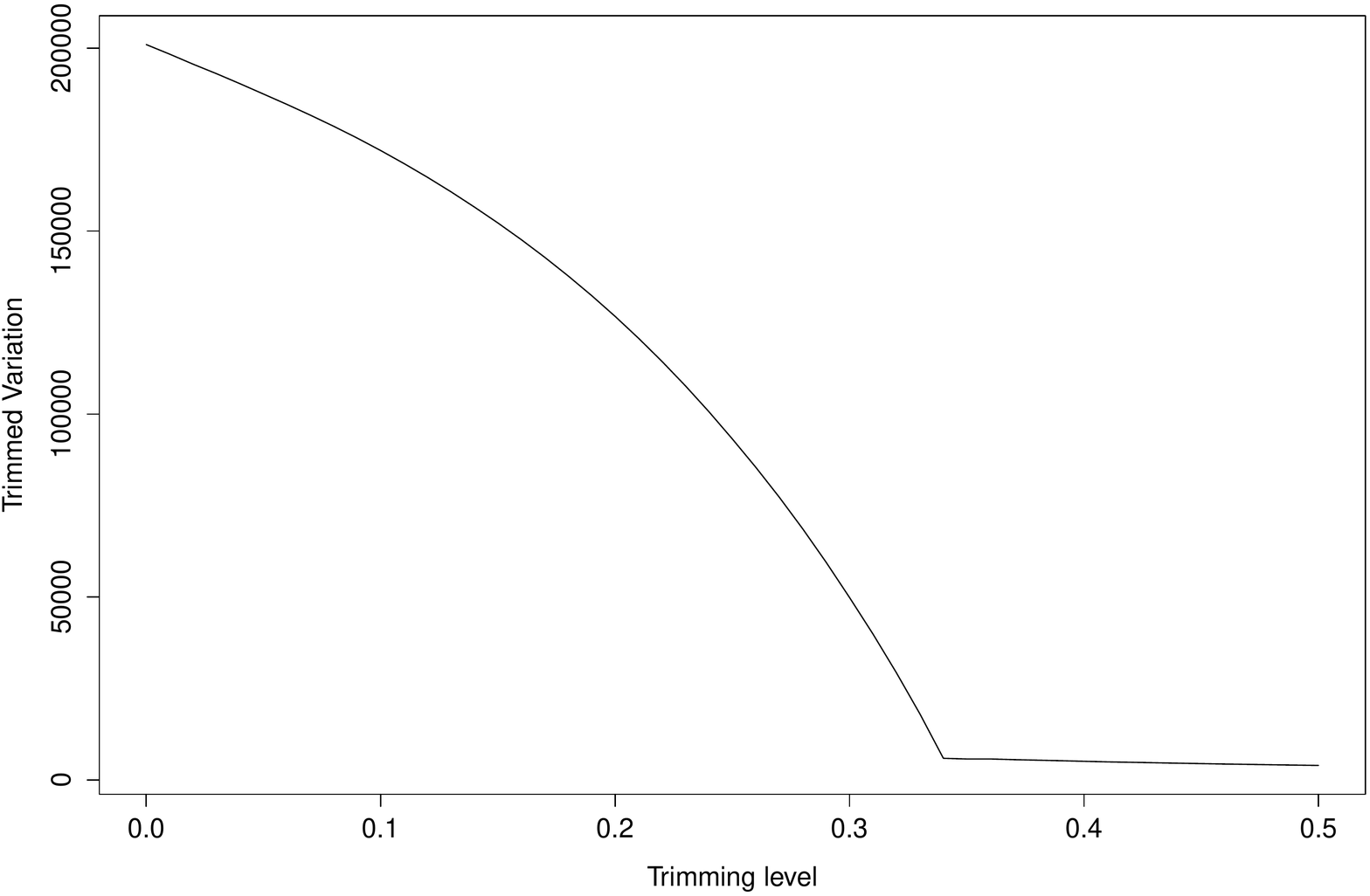}
\vspace{-.5cm}
\caption{Plot showing the evolution of the trimmed variations vs trimming levels on the 100 MCD's solutions in the Palomar Data example. }
\label{palomar2}
\end{center}
\end{figure}

\begin{Ejem}{\rm
Let us assume that we are carrying out an experiment in $k=100$ hospitals on a 2-dimensional r.v., that we are taking a sample with size $n=100$ in each hospital, and that each hospital is  sending only its own estimation of the mean and the covariance matrix based on the sample in its study. 

Let us also assume that the population is divided in two subpopulations. The first subpopulation is composed by 90\% of individuals and  the distribution of the variable of interest in this subpopulation is  standard normal, while the distribution in the second subpopulation is also normal, with the identity as covariance matrix and the mean at $(4,4)$. The real goal of the study is the estimation of the parameters of the majority, the second subpopulation being considered as composed by outliers.  

The statistician in charge of the experiment,  being aware of these issues, decides that each hospital uses the Minimum Covariance Determinant method (MCD, proposed in Rousseeuw \cite{RousseeuwMCD}), based on 80\% of the points in its sample to estimate the mean and covariance matrix of the people in its area (similar results could be obtained through the procedure developed in Cuesta-Albertos et al \cite{Cuesta-Albertos2008}), the reason to choose these estimators being that the probability of obtaining more than 20 outliers in a binomial sample with parameters $n=100$ and $p=.1$ being 0.00081 and, as long as we obtain less than 20 outliers in a sample with size 100, the MCD method will give a fair estimation of the parameters in the main subpopulation.

However, it happens that, unknown to the statistician, the population is relatively heterogeneous, and that, in fact, the proportion of people in a given area belonging to the second subpopulation is chosen using a distribution Beta with parameters (4,36), which gives a global proportion of 0.1, but  irregularly scattered. 

We have made a simulation of this process resulting that 5 hospitals have got more than 20 outliers, leading to largely  wrong estimations of the parameters. The results of this experiment appear in the left hand side graph in Figure \ref{Elipses_3}. There, most estimations appear in grey, but a few of them have been drawn in black to give a general idea of the objects we have obtained in the first part of the process.

\begin{figure}[tb] 
\begin{center}
\includegraphics[width=7cm]{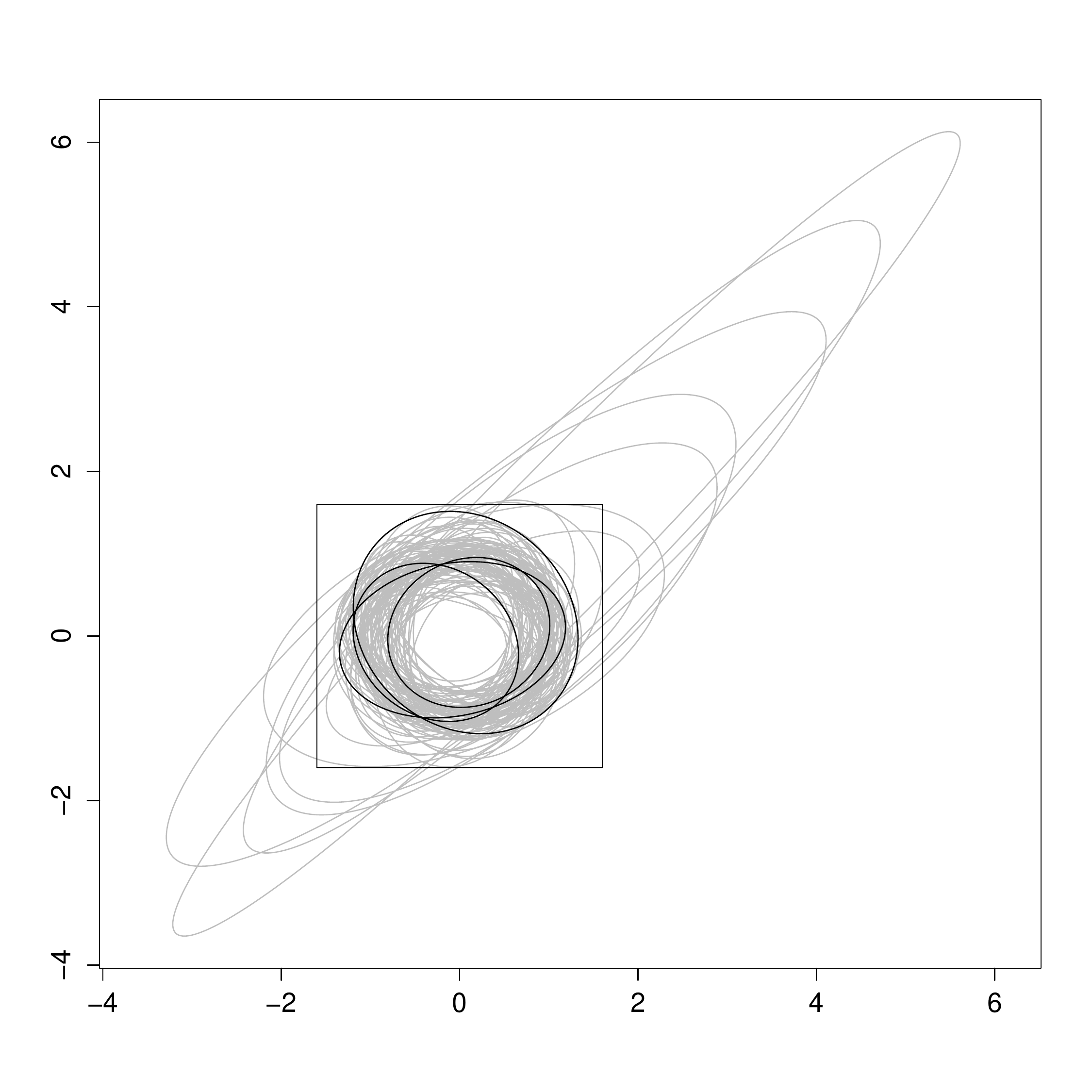}
\includegraphics[width=7cm]{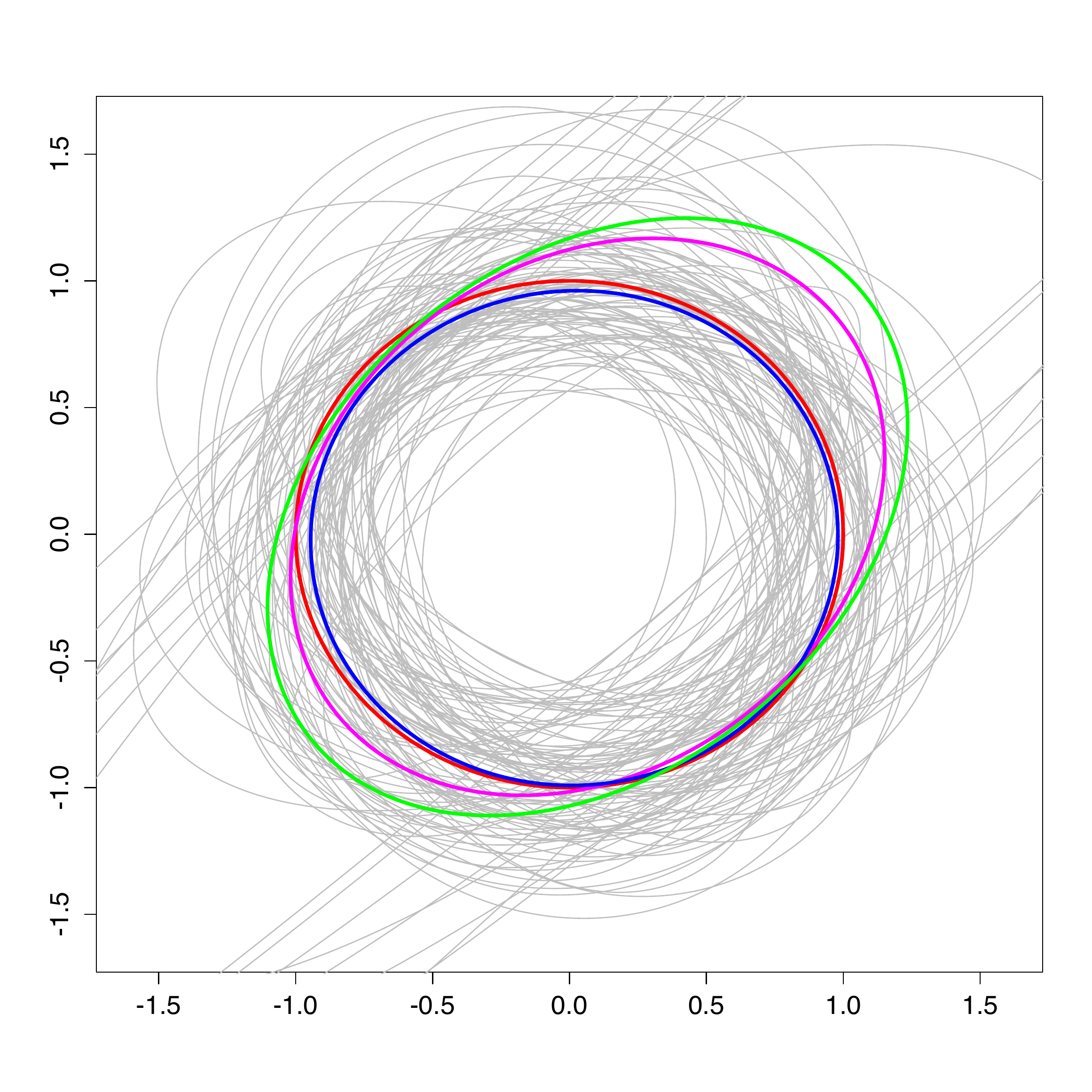}
\vspace{-.5cm}
\caption{Computation of trimmed barycenters (right figure) of the ellipses shown in the figure in the left}
\label{Elipses_3}
\end{center}
\end{figure}

The right hand side graph presents the area inside the square in the left hand side graph with some summarizing possibilities for the estimations shown in the left graph. Here the red ellipse represents the standard normal distribution (which can be considered as our target since this distribution produced most of the data in the analyzed samples). The green ellipse represents the normal distribution whose mean (resp. covariance matrix) is the sample mean of the estimated means (resp. covariance matrices). This estimator is not expected to be particularly good.

The magenta ellipse represents the (non-trimmed) barycenter. This estimation is  affected by the anomalous estimations (but less that the previous one). 
The blue ellipse represents the $0.2$-trimmed barycenter which, practically, matches the target.
}
\end{Ejem}
\begin{Ejem}\label{palomardata}{\rm
The Palomar Data is a data set considered in Rousseeuw and Driessen \cite{RousseeuwDriessen}, consisting in astronomic measurements recorded at the California Institute of Technology within the Digitized Palomar Sky Survey. The set handled here, kindly shared by the authors, is the same analyzed in that paper, containing 132,402 observations in 6 variables. The analysis there showed the interest of considering robust estimations of the covariance matrix and related metrics instead of the crude Mahalanobis distance, obtained through the sample covariance matrix. In fact, through a plot of MCD-based robust Mahalanobis distances, they found evidence on the existence of several groups in the data and, as a key part of the fast MCD algorithm for large data sets, introduced a pooling strategy on the initial subsets of the data leading to the better solutions. Our approach looks for the comparison between the MCD solution achieved for the whole data set and those provided from 100 randomly chosen subsamples of size 5,000. Figure \ref{palomar} is a plot on the two first variables (MAperF and csfF)  of the data. It shows (gray) the 100 ellipses associated to the MCD's based on subsamples, that of the MCD based on the full sample (black dashed). It also includes the ellipses that result from several aggregations of the MCD's produced by the subsamples. The green one is just that associated to the mean of the 100 covariance matrices and centered in the mean of the 100 means estimations. In black, red, blue and magenta are  represented the trimmed barycenters of the 100 MCD's respectively corresponding to the trimming levels $\alpha=.1, .2, .3, .4.$ Figure \ref{palomar2} is the plot of trimmed variations vs trimming levels associated to the 100 MCD's solutions.

Through these pictures we have a nice summary. From both figures it becomes apparent that nearly  35\% of the solutions correspond to ellipses centered  around (18500,1000) with little variation within this group, while the remaining 65\% are very similar to the MCD obtained with the complete sample. This implies that the right solutions should be selected when trimming, at least, that (35\%) proportion. In agreement with the conclusions of the analysis carried in \cite{RousseeuwDriessen}, such behavior would suggest the existence of at least two main bulks of data. Although most samples have a proportion of data coming from these bulks that justify the MCD based on the complete sample, small variations in these proportions would consistently produce a very different MCD. In this situation, aggregation methods based on simple average would typically produce bad solutions, while monitoring the trimmed barycenter solutions allows a  well-founded, stable, ``wide consensus"  proposal.

To give evidence of feasibility of the proposal, we give below some details on the execution times of the involved procedures. Computations have been carried on a MacBook Pro with a 4 Ghz processor Intel Core i7 and 16 Gb of RAM. The MCD's have been computed with TCLUST (available at the CRAN, see Fritz et al \cite{Fritz}), an R application for model based robust clustering. The  parameters for the solution based on the full sample were k=1, alpha = 0.5, nstart = 150, restr.fact = 1e10, iter.max = 200, equal.weights = F. The only change in these parameters for the subsamples was  iter.max that was set to 100. The computations of trimmed barycenters have been also carried into the R framework, with programs based on the algorithm presented in this section.

\indent
Runtimes in seconds: For the large MCD (sample of size 132,402) 120.497 sec; for 100 MCD's (on samples of size 5000) 45.125 sec; for the .3-trimmed barycenter of the 100 MCD's solutions 30.985 sec; for the (51) $\alpha$-trimmed barycenters and trimmed variances (to produce the plot on the right in Figure \ref{palomar}, $\alpha=k/100$ for $k=0,\dots,10$) 1744.315 sec.
Handling the MCD based on the complete sample as reference, the squares of the Wasserstein distances to the  average solution and to those given by the trimmed barycenters for .1, .2, .3, .4 were respectively: 87260.07, 71459.66, 33953.8, 6426.18, 357.25.

Repeating the whole process under the same conditions, but with  subsample sizes of 10000 instead of 5000, the only runtime that changed was the corresponding to the 100 MCD's (on samples of size 10000) 110.007 sec. The squares of the distances were now: 73850.38, 54857.57, 21578.75, 1517.071, 175.90.}
\end{Ejem}

\section{Technical details and proofs}\label{appendix}

\subsection{{ Supplementary }results on Wasserstein spaces}\label{WassSpaces}

{
For ease of reference, we include in this section some relevant results on Wasserstein spaces for reference through the work.
From a technical point of view a great deal of interest on the Wasserstein distance $\Wd$ comes from the fact that 
it metrizes the weak convergence of probabilities plus the convergence of their second order moments: Given $(P_n)_n \subset \Pd$ and $P \in \Pd$,
\begin{equation}\label{convergences}
\Wd(P_n,P) \conv 0 \ \ \mbox{ if and only if } \ P_n \convw P \ \ \mbox{ and } \ \int_{\Read}\|x\|^2P_n(dx) \conv \int_{\Read}\|x\|^2P(dx).
\end{equation}
More generally,  the following theorem gives a  very useful characterization  
(see e.g. Theorem 7.12 in \cite{Villani2}) of the convergence in the space $W_2(\Pd)$. }

\begin{Theo}\label{convW2}
Let $(\mu_n)_n$ and $\mu$ be in $W_2(\Pd)$, and consider the probability degenerated at zero, $\delta_{\{0\}}$ (that can be substituted by any other fixed probability in $\Pd$). Convergence $\mathcal{W}_{\mathcal{P}_2}(\mu_n ,\mu)\conv 0$ holds if and only if:
\begin{equation}\label{condW2}
\mu_n \convw \mu \ \ \mbox{ and } \lim_{R\conv \infty} \limsup_{n\to \infty} \int_{\mathcal{W}_2(\delta_{\{0\}},P)>R}\Wdd(\delta_{\{0\}},P)\mu_n(dP)=0.
\end{equation}
\end{Theo}

\begin{Prop}\label{propiedades}
If the sequences $(\mu_n)_n, (\nu_n)_n$ in $W_2(\Pd)$, verify $\mu_n \convw \mu$ and $\nu_n \convw \nu$, then $\mathcal{W}_{\mathcal{P}_2}(\mu,\nu)\leq \liminf\mathcal{W}_{\mathcal{P}_2}(\mu_n,\nu_n)$. Moreover, if the convergences are in the  sense showed in (\ref{condW2}), then the convergence $\mathcal{W}_{\mathcal{P}_2}(\mu_n,\nu_n)\conv\mathcal{W}_{\mathcal{P}_2}(\mu,\nu)$ holds.
\end{Prop}

Note that  the uniform integrability condition in (\ref{condW2}) is similar to the uniform integrability condition of $\|x\|^2$  in (\ref{convergences}).

{ Existence and continuity properties of barycenters in $W_2(\Pd)$ are guaranteed by Proposition \ref{existuni} and Theorem \ref{consistencyLoubes}
to be stated next. They follow from the results in \cite{Le Gouic}.}

\begin{Prop}\label{existuni}
If $\mu \in W_2(\Pd)$ and every $P$ in the support of $\mu$ is absolutely continuous, then the barycenter of the random measure $\mu$ exists and it is unique. 
\end{Prop}

\begin{Theo}\label{consistencyLoubes}
Let $\left( \mu_j \right)_{j=1}^\infty \subset W_2(\Pd)$ and
set $\bar \mu_j$ a barycenter of $\mu_j$, for all $j=1,\dots$ Suppose that for some $\mu \in W_2(\Pd)$ we have
that $\mathcal{W}_{\mathcal{P}_2}(\mu, \mu_j)\to 0.$ Then, the sequence $\left( \bar \mu_j \right)_{j=1}^\infty$ is precompact in $\Pd$ and any limit
is a barycenter of $\mu$.
\end{Theo}

In particular, when the limit distribution $\mu$ has only one barycenter, this theorem ensures convergence in $\Pd$ of  the barycenters to that of $\mu$. In a sample setting, when the probability measures $\mu_n$ are the sample ones  giving weight $1/n$ to the first $n$ probabilities $P_1,\ldots,P_n$ obtained as realizations of the random probability measure $\mu \in W_2(\Pd),$ by Varadarajan theorem, $\mu_n \convw \mu$ almost surely. Now let us consider the probability degenerated at zero, $\delta_{\{0\}}.$ Since the classical Strong Law of Large Numbers applied to the real i.i.d. random variables $\Wdd(P_i, \delta_{\{0\}})$ gives 
$$
\int_{\Pd}\Wdd(P, \delta_{\{0\}})\mu_n(dP)=\frac 1 n \sum_{i=1}^n\Wdd(P_i, \delta_{\{0\}})\convs \int_{\Pd}\Wdd(P, \delta_{\{0\}})\mu(dP),
$$
the characterization in Theorem \ref{convW2} of convergence in the $\mathcal{W}_{\mathcal{P}_2}$ sense, through Theorem \ref{consistencyLoubes}, proves the following Strong Law of Large Numbers for barycenters. 

\begin{Theo}\label{consistency}
Assume that $\mu \in W_2(\Pd)$ and {that the barycenter of $\mu$ is unique}. If $\mu_n$ is the sample probability giving mass $1/n$ to the probabilities $P_1,\ldots,P_n$ obtained as independent realizations of $\mu$, then the barycenters are consistent, i.e. $\Wd(\bar\mu_n, \bar\mu)\convs 0.$
\end{Theo}

\subsection{Overview on trimming}\label{overview}

In this section we recall some important properties of probability trimmings and obtain new results of interest in our current framework. In particular we emphasize those connected with Wasserstein spaces and distances. We begin providing a list of statements arising from \cite{Alvarez-Esteban2011}, that can be easily translated to the framework of Polish spaces (metri\-zable, separable and complete spaces). 

\begin{Prop}\label{equivalencias}
Let $P$ be a probability in any mesurable space $(\Omega,\sigma)$ and $\alpha \in [0,1)$. The following statements are equivalent:
\begin{itemize}
\item[{\em (a)}] The probability $P^*$ is a trimmed version of $P$. 
\item[{\em (b)}] 
$P^*$ is absolutely continuous with respect to  $P$, and $\frac{dP^*}{dP}\leq \frac{1}{1-\alpha}$
\item[{\em (c)}] 
$(1-\alpha)P^*(A)\leq P(A)$ for every  set $A \in \sigma.$
\end{itemize}
\end{Prop}

\begin{Prop}\label{newproposition2}
Let $P$ be a probability in any abstract space and $T$ a measurable map taking values in a Polish space. If  $T$ transports $P$ to $Q$,  then for every $\alpha$
$$\mathcal{T}_{\alpha}(Q)= \left\lbrace  P^*\circ T^{-1}: \, P^*\in\mathcal{T}_\alpha(P) \right\rbrace. $$
\end{Prop}

\begin{Prop}\label{proposition1} Let $(E,d)$ be a Polish space and $\alpha\in (0,1)$.

\begin{itemize}

\item[{\em (a)}] 
If  $P$ is any probability measure on $(E,d)$, then $\mathcal{T}_\alpha(P)$ is compact for the topology of weak convergence.
\item[{\em (b)}]
 If $\left\lbrace P_{n}\right\rbrace_n$   is a tight sequence of probabilities  on $(E,d)$ and $P^*_{n} \in \mathcal{T}_{\alpha}(P_n)$ for every $n$, then $\left\lbrace P^*_{n} \right\rbrace_n$ is tight. Moreover, if $P_{n} \to_{w} P$ and   $P^*_{n}  \to_{w} P^*$, then $P^* \in \mathcal{T}_{\alpha}(P)$.
\end{itemize}
\end{Prop}

\begin{Prop}\label{compact}
If $0<\alpha <1$ and $P\in {W}_2(\Pd)$ or $P \in \Pd$,  then $\mathcal{T}_\alpha
(P)$ is compact in the $\mathcal{W}_2$ topology.
\end{Prop}
\noindent{\bf Proof:}  The proof given in \cite{Alvarez-Esteban2011} for $P \in \Pd$ quickly extends to the { case} $P\in {W}_2(\Pd)$ by handling  the uniformly integrability condition in Theorem \ref{convW2}.
\FIN

\begin{Prop} \label{lema} 
Let $0<\alpha <1$, $\left\lbrace P_{n}\right\rbrace_n$ and $P$ be probabilities on a Polish space $(E,d)$, and assume that $P_{n} \to_{w} P$. 
Then, if $P^* \in  \mathcal{T}_{\alpha}(P)$, there exists a sequence   $\left\lbrace P^*_{n}\right\rbrace_n$ such that  
$P^*_{n} \in  \mathcal{T}_{\alpha}(P_n)$, for all $n$, and $P^*_{n}\to_{w} P^*$.
\end{Prop}
\noindent{\bf Proof:} Use Skorohod's Representation Theorem (see e.g. Theorem 11.7.2 in Dudley \cite{Dudley}), to obtain $E$-valued 
measurable maps $X,X_1,\ldots$ defined on a probability space $(\Omega,\sigma,\Prob)$ such that $\mathcal{L}(X_n)=P_n$, 
$\mathcal{L}(X)=P$, and $X_n \conv X$, $\Prob-$a.s. 

By Proposition \ref{newproposition2}, $P^* \in  \mathcal{T}_{\alpha}(P)$ can be represented as $P^*=\Prob^* \circ X^{-1}$ for some $\Prob^*\in \mathcal{T}_{\alpha}(\Prob)$. By considering $P_n^*:=\Prob^* \circ X_n^{-1}$, we obtain probabilities in $\mathcal{T}_{\alpha}(P_n)$, that obviously converge weakly to $P^*$ because $X_n \conv X$ also $\Prob^*-$ a.s.
\FIN

\begin{Nota}\label{remark}
{\rm
Note that any kind of uniform integrability condition like the one in (\ref{condW2}) verified for some sequence $\left\lbrace P_{n}\right\rbrace_n$ is automatically shared for any sequence $\left\lbrace P_{n}^*\right\rbrace_n$ such that $P_{n}^*\in \mathcal{T}_{\alpha}(P_n)$ for every $n$. Therefore Proposition \ref{lema} and (\ref{condW2}) imply that if $P_n \conv P$ in $\mathcal{W}_{\mathcal{P}_2}$, then the sequence $\left\lbrace P_{n}^*\right\rbrace_n$ is precompact in $\mathcal{W}_{\mathcal{P}_2}$ and any limit belongs to $\mathcal{T}_{\alpha}(P)$.
}
\end{Nota}

\subsection{Proofs of Propositions \ref{equiv} and \ref{standardDev}} \label{Sec.ProofProposition}

\textbf{Proof of Proposition \ref{equiv}:}
A similarity transformation, $T$, can be expressed   as a linear transformation $T=cA+b$, 
where $c$ is a constant, $A$ an orthogonal transformation and $b\in\Read$. 
If $(X,Y)$ is an $\Wd$-o.t.p. for the probabilities $(P,Q)$, and $(AX^*,AY^*)$ is an $\Wd$-o.t.p. for $(P\circ A^{-1},Q\circ A^{-1})$ then we have 
\begin{eqnarray*}
\Wdd(P,Q)&=&E\|X-Y\|^2=E\|AX-AY\|^2\geq \Wdd(P\circ A^{-1},Q\circ A^{-1})
\\ 
&=&E\|AX^*-AY^*\|^2=E\|X^*-Y^*\|^2\geq\Wdd(P,Q),
\end{eqnarray*}
hence $\Wdd(P\circ A^{-1},Q\circ A^{-1})=\Wdd(P,Q)$, and for $T$ we easily obtain $\Wdd(P\circ T^{-1},Q\circ T^{-1})=c^2\Wdd(P,Q)$. Therefore, for every $Q$,
we have
\begin{eqnarray*}\label{equi}
 \int_{\Omega}\mathcal{W}_2^2(\mu_\omega\circ T^{-1},\bar \mu \circ T^{-1})\Prob(d\omega) 
 &=& c^2 \int_{\Omega}\mathcal{W}_2^2(\mu_\omega,\bar \mu)\Prob(d\omega) 
 \\
 &\leq& c^2\int_{\Omega}\mathcal{W}_2^2(\mu_\omega,Q)\Prob(d\omega)
 \\
 &= &
 \int_{\Omega}\mathcal{W}_2^2(\mu_\omega\circ T^{-1},Q \circ T^{-1})\Prob(d\omega).
 \end{eqnarray*}
 Since $T$ is invertible, denoting $S=T^{-1}$, every $Q$ can be written as $Q=Q^*\circ S^{-1}$ for some $Q^*$, hence we deduce that
 $$\int_{\Omega}\mathcal{W}_2^2(\mu_\omega\circ T^{-1},\bar \mu \circ T^{-1})\Prob(d\omega)  \leq \int_{\Omega}\mathcal{W}_2^2(\mu_\omega\circ T^{-1},Q^* )\Prob(d\omega) \mbox{ for every } Q^*\in \Pd.$$
\FIN
 
 \medskip
\noindent
\textbf{Proof of Proposition \ref{standardDev}:}
Let $X$ be a random vector with $\mathcal L(X)=\bar P,$ and $T_j, j=1,\dots,k$ be optimal transport maps for $(\bar P,P_j)$. Denoting $X_j=T_j(X)$, we know that $\mathcal L(X_j)=P_j$ but also, by  Proposition \ref{basicineq}, $\bar P=\mathcal L(\sum_{j=1}^k\lambda_jX_j)$. Therefore, by Minkowski inequality, we have
$$(E\|X\|^2)^{1/2}=(E\|\sum_{j=1}^k\lambda_jX_j\|^2)^{1/2}\leq \sum_{j=1}^k(E\|\lambda_jX_j\|^2)^{1/2}=\sum_{j=1}^k\lambda_j(E\|X_j\|^2)^{1/2}.$$

\FIN

\subsection{Existence and consistency of the trimmed barycenter} \label{Sec.Consistency}

 Let us begin noting that, under the additional assumption  $\mu \in W_2(\Pd)$, the results would easily follow from Theorem \ref{consistencyLoubes} and the compactness of the set $\mathcal{T}_{\alpha}(\mu)$ stated in Proposition \ref{compact}. However, as stated in Theorem \ref{Theo.Bary}, that assumption is not needed at all.

\medskip
\noindent
\textbf{Proof of Theorem \ref{Theo.Bary}:} 
Recall definition (\ref{def1trimbar}) and assume that $\mu_n^*\in \mathcal{T}_{\alpha}(\mu)$ and $\nu_n \in \Pd$ verifying
\begin{equation}\label{minimizing}
\int \Wdd(P,\nu_n)\mu_n^*(dP) \conv \mbox{Var}_\alpha(\mu).
\end{equation}
 We already know that $\mbox{Var}_\alpha(\mu)$ is finite and that we can assume that every  $\mu_n^*$ in the the minimizing sequence belongs to $W_2(\Pd)$, hence the $\nu_n$'s can be chosen as their barycenters. Thus, we will take $\nu_n=\bar \mu_n^*$.

The next step is to show that the sequence $\{\int \|x\|^2 \bar \mu_n^* (dx)\}_n$ as well as that of their associated radii   $\{r_\alpha(\bar \mu_n^*)\}_n$ (defined in (\ref{radios})) are bounded.  { For the sake of readability}, we state this result as  a lemma.

\begin{Lemm} \label{Lemm.CotaRadios}
Let $\{ Z_n\}_n$ be a sequence of r.v.'s defined on some probability space $(\Omega,\sigma,\Prob)$ such that $\Prob_{Z_n}= \bar \mu_n^*$ for every \nin . Then, it happens that $M:=\sup_n \Exp[\|Z_n\|^2]< \infty$. Moreover, the sequence $\{r_\alpha(\bar \mu_n^*)\}_n$ is bounded.
\end{Lemm}
\medskip
\noindent
\textbf{Proof:}
Take $r_0>0$ such that
$
p:= \mu [B_{\mathcal{W}}(\delta_{\{0\}},r_0)] > \alpha.
$
Let us assume that there exists a subsequence such that $\lim_n \Exp(\|Z_{k_n}\|^2) = \infty$. For this subsequence, we  have that if $P \in B_{\mathcal{W}}(\delta_{\{0\}},r_0)$, then, since \Wd \ is a metric,
\[
\Wd (P, \bar \mu^*_{k_n}) \geq \Wd ( \delta_{\{0\}} , \bar \mu^*_{k_n}) - \Wd(P,\delta_{\{0\}} ) 
\geq
(\Exp(\|Z_{k_n}\|^2))^{1/2} - r_0,
\]
and, consequently, since $p > \alpha$,
\[
\int \Wdd ( P , \bar \mu^*_{k_n}) \mu_{k_n}^* (dP) \geq 
\left( (\Exp(\|Z_{k_n}\|^2))^{1/2} - r_0 \right)^2 ( p - \alpha)
\to \infty,
\]
which contradicts the minimizing property (\ref{minimizing}) of the chosen sequence  with $\mbox{Var}_\alpha(\mu)<\infty$. Thus, the sequence $\{\Exp [\|Z_n\|^2]\}_n$ is bounded.
Now, if $P \in B_{\mathcal{W}}(\delta_{\{0\}},r_\alpha(\delta_{\{0\}}))$, then
\[
\Wd (P, \bar \mu_n^*) \leq \Wd (P, \delta_{\{0\}}) + \Wd (\delta_{\{0\}},\bar \mu_n^*) \leq r_\alpha(\delta_{\{0\}}) + (E(\|Z_n\|^2))^{1/2},
\]
which implies that the set $B_{\mathcal{W}}(\delta_{\{0\}},r_\alpha(\delta_{\{0\}})) $ is a subset of the ball with center at $\bar \mu_n^*$ and radius $ r_\alpha(\delta_{\{0\}}) + (E(\|Z_n\|^2))^{1/2}$, and therefore, $r_\alpha(\bar \mu_n^*) \leq 
 r_\alpha(\delta_{\{0\}}) + \sup_m (E(\|Z_m\|^2))^{1/2} $ for every \nin.
\FIN

Returning to the proof  of Theorem \ref{Theo.Bary}, note that, by the first result of the lemma, $\{\bar \mu_n^*\}_n$ is tight, so w.l.o.g. we can assume that it converges in distribution to some $\nu_0\in \Pd$. Moreover, by the lemma, the supports of the associated trimmed probabilities $\mu_n^*$ are contained in a common ball $B_{\mathcal{W}}(\delta_{\{0\}},M+\sup\{r_\alpha(\bar \mu_n^*), \nin \})$ in $\Pd$, thus  we can also assume that it converges to some $\mu_0^*\in \mathcal{T}_{\alpha}(\mu)$ weakly and (by uniform integrability) in \Wd. This implies, by Theorem \ref{consistencyLoubes}, that the limit $\nu_0$ of the barycenters must be a barycenter of $\mu_0^*$ and that the convergence is also in \Wd. 

By continuity of \Wd \ we have  $\mathcal{W}_{\mathcal{P}_2}(\mu_n^*,\delta_{\{\bar \mu_n^*\}})\conv\mathcal{W}_{\mathcal{P}_2}(\mu_0^*,\delta_{\{\nu_0\}})$, leading also to
\[
\lim_n \int \Wdd(P,\bar \mu_n^*)\mu_n^*(dP)=\lim_n\Wdd(\mu_n^*,\delta_{\{\bar \mu_n^*\}}) =\Wdd(\mu_0^*,\delta_{\{\nu_0\}})=\int \Wdd(P,\nu_0)\mu_0^*(dP),
\]
that shows that $\nu_0$ is a trimmed barycenter of $\mu$.
\FIN

An easy modification of this proof allows to guarantee a consistency result in the sense of Theorem \ref{consistencyLoubes}  also without the integrability assumption.

\medskip
\noindent
\textbf{Proof of Theorem \ref{consistencytypeLoubes}:}

\noindent
Since $\mu_n \convw \mu$, we can choose a large enough $M>0$  such that $\mu_n[B_{\mathcal{W}}(\delta_{\{0\}},M)]>1-\alpha$ for every \nin. This implies that there exist trimmed versions $\mu_n^*\in \mathcal{T}_{\alpha}(\mu_n)$ with support contained in $B_{\mathcal{W}}(\delta_{\{0\}},M)$. Therefore we have that
$
\mbox{Var}_\alpha(\mu_n)\leq \mathcal{W}_{\mathcal{P}_2}(\mu_n^*,\delta_{\{\delta_{\{0\}}\}})\leq M,
$
and $\limsup_n \mbox{Var}_\alpha(\mu_n)\leq M<\infty.$

From this point we can repeat the proof of Lemma \ref{Lemm.CotaRadios} to guarantee that the sequence of trimmed barycenters is contained in a large enough ball $B_{\mathcal{W}}(\delta_{\{0\}},M)$ and that the sequence of associated radii $(r_\alpha(\bar \mu_n^*))_n$ is bounded. The argument at the end of the proof of Theorem \ref{Theo.Bary} applies also here to prove that weakly convergent subsequences of trimmed barycenters must converge also in $\Wd$ and that the limits must be trimmed barycenters of the limit law $\mu.$
\FIN

\subsection{Proofs of Theorems  \ref{extensionGelbrich} and \ref{self-suf}}\label{locscale}

Recall that $\mathcal{F}(P_0):=\{\mathcal{L}(A{\bf X}_0+m): A\in {\cal M}_{d\times d}^+, m\in\Read\}$ is the location-scatter family  induced by  positive definite affine transformations from the law $P_0=\mathcal L ({\bf X}_0)$. We assume throughout that $P_0$ is absolutely continuous as an easy way to guarantee uniqueness of optimal transport maps and of barycenters, but much of the following analysis does not depend of this assumption.  As we already noted, we can assume w.l.o.g. that   $P_0$ has zero mean and  covariance matrix  $I_d$. The probabilities in $\mathcal{F}(P_0)$ are represented as $\mathbb{P}_{m,\Sigma}$, where $m$ is the mean, and  $\Sigma$ the covariance matrix of the probability under consideration.

Relation (\ref{distancelocscale})  allows to extend Theorem \ref{casonormal} to any family $\mathcal{F}(P_0)$ in a simple way. However we will give a direct proof.
 For this task  let us include  the following proposition already obtained in Cuesta-Albertos et al \cite{Cues00}. It will allow us to guarantee that barycenters of families of absolutely continuous probabilities in \Pd  cannot be degenerated on subspaces of dimension lower than $d$.

\begin{Prop}\label{nondegeneracy}
Let $P,Q \in \Pd$. Let us assume that $P \in \mathcal{P}_{2,ac}$  and that $Q$ is supported on the  subspace generated by the first $q$ components of $\Rea^d$, with $q < d$. Denote by $T^{1,\ldots,q}$ the \Wd \ optimal map transporting the marginal probability, $P^{1,\ldots,q}$, of $P$ on that subspace to $Q$.
Then the map $T(x_1,\ldots,x_d):= T^{1,\ldots,q}(x_1,\ldots,x_q)$ is a $\Wd$ optimal map  transporting $P$ to $Q$. 
\end{Prop}

\begin{Prop}\label{abscontbarycenter}
Let $\mu \in W_2(\Pd)$ and, using  the notation employed in (\ref{notacionbar}), assume that for every $\omega\in\Omega,$ the probability $\mu_\omega$ is absolutely continuous. Then, the barycenter  of $\mu$ cannot be supported on an affine subspace of dimension $q<d$.
\end{Prop}

\noindent
{\bf Proof.}
 Let $\mu \in W_2(\Pd)$, such that $\mu_\omega$ is absolutely continuous for every $\omega\in\Omega$ and let $m_\omega$ be the mean of $\mu_\omega$. Under these conditions, existence and uniqueness of the barycenter  are guaranteed by Proposition \ref{existuni}. Since it is trivial to show that the mean of the barycenter coincides with $\int_\Omega m_\omega \Prob(d\omega)$, we can simplify the problem by considering centered in mean distributions (that is, $m_\omega=0$ for every $\omega$) which remain absolutely continuous.  Let $\bar \mu$ be the  barycenter  (with zero mean) of $\mu$, so suppose that it is supported on a subspace (instead of a general affine subspace) of dimension $q<d$. We can assume, w.l.o.g., that $\bar \mu$ is supported on the subspace corresponding to the first $q$ components. Let $\mu_\omega^{1,\ldots,q}$ denote the marginal of $\mu_\omega$ on this subspace. Since $\mu_\omega^{1,\ldots,q}\ll \ell_q,$ we know that there exists an optimal map $T_\omega^{1,\ldots,q}$ transporting $\mu_\omega^{1,\ldots,q}$ to $\bar \mu$. From the previous proposition the map $T_\omega$ defined by $T_\omega(x_1,\ldots,x_d):=T_\omega^{1,\ldots,q}(x_1,\ldots,x_q)$ is an optimal map transporting $\mu_\omega$ to $\bar \mu.$ Therefore we have 
\begin{equation}\label{Eq.marginales}
 \Wdd(\mu_\omega,\bar \mu)=\Wdd(\mu_\omega^{1,\ldots,q},\bar \mu) + \sum_{j=q+1}^d\Wdd(\mu_\omega^j,\delta_{\{0\}}),
 \end{equation}
where $\mu_\omega^j$ is the $j-$th marginal of $\mu_\omega$.

Let us consider the probability $ \mu^*:=\bar \mu \times \bar \mu^{q+1} \times \dots \times \bar \mu^d$ and denote by $\bar \mu^j$  the barycenter of the probability $\mu^j\in W_2(\mathcal{P}(\mathbb{R}))$, which is not degenerated because  $\mu_\omega^j \ll\ell_1$ for every $j$ (recall the comments preceding Theorem \ref{casonormal}). Thus, from (\ref{Eq.marginales}), we have
\begin{eqnarray*}
\int_\Omega\Wdd(\mu_\omega,\bar \mu)\Prob(d\omega)
&=&
\int_\Omega\Wdd(\mu_\omega^{1,\ldots,q},\bar \mu)\Prob(d\omega)+ \sum_{j=q+1}^d\int \Wdd(\mu_\omega^j,\delta_{\{0\}})\Prob(d\omega) 
\\ 
&>&
\int_\Omega\Wdd(\mu_\omega^{1,\ldots,q},\bar \mu)\Prob(d\omega)
+ 
\sum_{j=q+1}^d\int_\Omega \Wdd(\mu_\omega^j,\bar \mu^j)\Prob(d\omega)
\\
&=&  \int_\Omega\Wdd(\mu_\omega, \mu^*)\Prob(d\omega),
\end{eqnarray*}
contradicting the character of barycenter of $\bar \mu$.
\FIN

\noindent
\textbf{Proof of Theorem \ref{extensionGelbrich}:}

\noindent
Let $P \in \Pd$ and let $N$ be a normal law with the same mean and covariance matrix as $P$. From Gelbrich's bound (\ref{cotaGelbrich}), we have $\Wdd(P_i, P)\geq\Wdd(N_i,N)$ for $i=1,\ldots,k$, hence 
\begin{equation} \label{desig1}
\sum_{i=1}^k\lambda_i \Wdd(P_i, P)\geq \sum_{i=1}^k\lambda_i \Wdd(N_i,N)\geq \sum_{i=1}^k\lambda_i \Wdd(N_i,\bar N).
\end{equation}
 Moreover, according to Theorem \ref{Gelbrich2}, equality in the first inequality is only possible if $P_i \in \mathcal{F}(P), i=1,\ldots,k$.
On the other hand, let $\mathbb{P}^*$ be the probability law in  $\mathcal{F}(P_0)$ with the same mean  and covariance matrix as the barycenter $\bar N$ of $\{N_i\}_{i=1}^k$. Then we have 
$$
 \sum_{i=1}^k\lambda_i \Wdd(N_i,\bar N)=\sum_{i=1}^k\lambda_i \Wdd(\mathbb{P}_{m_i,\Sigma_i}, \mathbb{P}^*)\geq  \sum_{i=1}^k\lambda_i \Wdd(\mathbb{P}_{m_i,\Sigma_i},\bar {\mathbb{P}}).
 $$

Particularizing the first inequality in (\ref{desig1}) for $P_i=\mathbb{P}_{m_i,\Sigma_i}, i=1,\ldots,k$ and $P=\bar {\mathbb{P}}$, the concatenation with the last chain of inequalities gives that a normal law with the same mean and covariance matrix as $\bar {\mathbb{P}}$ would be a barycenter for $\{N_i\}_{i=1}^k$. The uniqueness of this barycenter implies that $\bar {\mathbb{P}}$ and $\bar N$  must have the same mean and covariance matrix.

The proof ends by considering $P=\bar P$ in (\ref{desig1}) because both equalities would imply that the mean and the covariance matrix of $\bar P$ must coincide with those of $\bar N$ and also that $\bar P$ can be obtained from every $P_i$ through a positive definite transformation. By Proposition \ref{abscontbarycenter} these covariance matrices must be nonsingular, thus the barycenters, in particular $\bar P$, must be also absolutely continuous  and every  $P_i$ can be   obtained from $\bar P$ through a positive definite affine transformation, thus $\{P_i\}_{i=1}^k \subset \mathcal{F}({\bar P})$ holds.
\FIN

\medskip

The following lemma can be proved through elementary arguments (see, e.g., equation (18)  in \cite{preprint}) and will be used in the proof of uniqueness 
involved in Theorem \ref{self-suf}.

\begin{Lemm} \label{Lemm.ElUltimo}
Let $\Sigma_i, i=0, 1,2$ be positive definite matrices and define  $$\Sigma_{0,2}:=\Sigma_0^{-1/2}\left(\Sigma_0^{1/2}\Sigma_2\Sigma_0^{1/2}\right)\Sigma_0^{-1/2}.
$$

Let $X_i, i=1,2$ be random vectors on \Read with nonsingular respective laws $\mathbb{P }_{0,\Sigma_i}  \in \mathcal{F}(P_0)$,  $i =1,2$. Then the inequality 
\[
\Wdd(N(0,\Sigma_1),N(0,\Sigma_2))\geq \mbox{trace}\left(\left(I_d-\Sigma_{0,2}\right)\Sigma_1\right)+\Exp\left(\|X_2\|^2-X_2^t\Sigma_{0,2}^{-1}X_2\right)
\]
holds. If $\Sigma_0=\Sigma_1$ and $X_2=\Sigma_{0,2}X_1$, then the inequality is an equality.
\end{Lemm}

\noindent
\textbf{Proof of Theorem \ref{self-suf}:}

\noindent
The statement about the mean of $\bar \mu$ is already known, thus let us simplify the problem assuming that every $\mu_\omega$ is centered in mean. By Proposition \ref{abscontbarycenter}, $\bar \mu$ must be absolutely continuous, hence its covariance matrix $\bar \Sigma$ must be nonsingular. To simplify the notation, let us denote $\bar P=\mathbb{P }_{0,\bar \Sigma}\in \mathcal{F}(P_0)$. From Gelbrich's bound we have
$$
\int\Wdd(\mu_\omega,\bar \mu)\Prob(d\omega)\geq \int\Wdd(\mu_\omega,\bar P)\Prob(d\omega),
$$
hence, by the uniqueness of the barycenter,  $\bar \mu= \bar P$, and $\bar \mu\in \mathcal{F}(P_0)$. If we consider the optimal maps $\bar T_\omega$ transporting $\bar \mu$ to $\mu_\omega$, and define $\bar T(x):=\int \bar T_\omega(x)\Prob(d\omega)$, we have
\begin{eqnarray*}
\int\Wdd(\bar \mu,\mu_\omega)\Prob(d\omega)
&=& \int\left(\int\|x-\bar T_\omega(x)\|^2\bar \mu(dx)\right)\Prob(d\omega)
\\
 &=&
  \int\left(\int\left(\|x-\bar T(x)\|^2+\|\bar T(x)-\bar T_\omega(x)\|^2\right) \Prob(d\omega)\right)\bar\mu(dx)
    \\
   &\geq&\int\left(\int\|\bar T(x)-\bar T_\omega(x)\|^2\bar\mu(dx) \right)\Prob(d\omega) \geq \int\Wdd(\bar \mu\circ\bar T^{-1},\mu_\omega)\Prob(d\omega)
\end{eqnarray*}
that (by the uniqueness) is possible only if $\bar \mu\circ\bar T^{-1}=\bar \mu$, i.e., if $\bar T(x)=x$ $\bar \mu-$a.s.

To finalize, observe that the optimal transport maps $\bar T_\omega$ from $\bar \mu$ to $\mu_\omega$, being probabi\-lities in $\mathcal{F}(P_0)$, take the form $\bar \Sigma^{-1/2}\left(\bar\Sigma^{1/2}\Sigma_\omega\bar\Sigma^{1/2}\right)^{1/2}\bar\Sigma^{-1/2}$ (see (\ref{transportenormales})), therefore (since $\bar \Sigma$ is positive definite) the relation $\bar T(x)=x$ $\bar \mu-$a.s. is equivalent to
$$\bar\Sigma=\int\left(\bar\Sigma^{1/2}\Sigma_\omega\bar\Sigma^{1/2}\right)^{1/2}\Prob(d\omega)$$
This proves that $\bar\Sigma$ verifies the integral equation. 
To prove that the integral equation has only a positive definite solution, let $\hat \Sigma$ be any positive definite matrix and define 
$$\Sigma_{0,\omega}:=\hat \Sigma^{-1/2}\left(\hat\Sigma^{1/2}\Sigma_\omega\hat\Sigma^{1/2}\right)^{1/2}\hat\Sigma^{-1/2} \mbox{ and } \hat\Sigma^*:=\int \Sigma_{0,\omega}\Prob(d\omega).
$$

If we apply Lemma \ref{Lemm.ElUltimo} first to $\Sigma_0=\hat \Sigma$, $\Sigma_1=\Sigma$ and $\Sigma_2=\Sigma_\omega$, later to $\Sigma_0=\Sigma_1=\hat \Sigma$ and $\Sigma_2=\Sigma_\omega$, subtracting the results and integrating, we have that
$$
\int\Wdd(\mathbb{P}_{0,\Sigma},\mathbb{P}_{0,\Sigma_\omega})\Prob(d\omega)-\int\Wdd(\mathbb{P}_{0,\hat\Sigma},\mathbb{P}_{0,\Sigma_\omega})\Prob(d\omega)\geq\mbox{trace}\left(\left(I_d-\hat\Sigma^*\right)\left(\Sigma-\hat\Sigma\right)\right).
$$

Thus, if $\hat \Sigma$ is a solution of the integral equation, we would have that  $\mathbb{P}_{0,\hat\Sigma}$ is the barycenter of $\mu$, and the uniqueness of the barycenter gives that $\bar \Sigma = \hat \Sigma$.
\FIN

 \end{document}